\documentclass[12pt]{article}
\usepackage{amssymb,graphicx}

\title{Shape Dynamics and Effective Field Theory}
\author{Tim A. Koslowski\\ \texttt{email: t.a.koslowski@gmail.com}\\ \it  Department of Mathematics and Statistics,\\ \it University of New Brunswick\\ \it Fredericton, NB, E3B 5A3,\\ \it Canada}

\begin{document}

\maketitle

\begin{abstract}
  Shape Dynamics is a gauge theory based on spatial diffeomorphism- and Weyl-invariance which is locally indistinguishable form classical General Relativity. If taken seriously, it suggests that the spacetime--geometry picture that underlies General Relativity can be replaced by a picture based on spatial conformal geometry. This classically well understood trading of gauge symmetries opens new conceptual avenues in many approaches to quantum gravity. I focus on the general implications for quantum gravity and effective field theory and consider the application of the Shape Dynamics picture in the exact renormalization group approaches to gravity, loop- and polymer- quantization approaches to gravity and low energy effective field theories. I also discuss the interpretation of known results through in the Shape Dynamics picture, in particular holographic renormalization and the problem of time in canonical quantum gravity.  
\end{abstract}

\newpage

\section{Introduction}

One of the most important open problems in theoretical physics is posed by the lack of a quantum theory of gravity. This lack has sparked a number of ambitious research programs that are aimed at finding a quantum field theory that contains general relativity as a classical limit. The vast majority of attempts to quantize gravity are rooted in ideas about spacetime geometry. This is motivated by the fact that classical General Relativity is a theory of spacetime geometry, which is expected to arise as an effective field theory of a more fundamental quantum theory of spacetime geometry. In fact, the Einstein--Hilbert Lagrangian consists precisely of the two lowest dimensional local operators $R\sqrt{|g|}$ and $\sqrt{|g|}$. These are precisely the operators that are, by standard dimensional arguments, expected to be relevant at low energies, making it very natural to think of General Relativity as a low energy effective field theory of quantum spacetime geometry.

The nice picture of General Relativity as an effective field theory is obscured by three puzzles. These are: (1) The dimensionless cosmological constant is unnaturally fine tuned to be small but non-vanishing. (2) Four-derivative operators are power-counting marginal in 3+1 dimensions. The inclusion of 4-derivative operators changes the structure of the theory by introducing additional degrees of freedom and some of these new degrees of freedom threaten unitarity. (3) The conformal factor of the metric appears to have a kinetic term with wrong sign. The perturbative non--renormizability of the Einstein-Hilbert action and these three puzzles are commonly understood as a hint that quantum gravity may require a rethinking of fundamental principles that shape our current understanding of quantum field theory and the nature of gravity.

The central idea of this contribution is based on the observation that the same physical system admits many distinct mathematical descriptions, each suggesting a different construction principle. In particular, in section \ref{sec:Minimal}, we see that sets of gauge symmetries can be traded for one another and that the physical interpretation of the dynamical degrees of freedom is subject to reinterpretation. Having this symmetry trading in mind, we consider in section \ref{sec:Ontology} two sets of construction principles for General Relativity: the familiar spacetime picture and the spatial conformal picture considered by Barbour and collaborators \cite{Barbour:1982gha,Barbour:2011dn,Anderson:2004bq,Anderson:2004wr,Barbour:2003gr,Barbour:2000qg}. We then consider in section \ref{sec:symmetryTrading} a general mechanism that allows to trade gauge symmetries \cite{Gomes:2011zi}. This is possible because the canonical generators of gauge transformations are first class constraints, which require gauge fixing 
conditions to result in definite evolution equations. If the gauge--fixing conditions can be chosen to be first class as well, then one can switch the role of gauge generators and gauge fixing conditions without changing any observable predictions of the system, because observables of a gauge theory are by definition equivalence classes of gauge--invariant phase space functions, where equivalence is defined through coinciding on the surface where the constraints hold. This switch allows one to trade one set of gauge transformations for another without changing physical predictions.

It turns out that this symmetry trading mechanism can be used to trade refoliation symmetry in the spacetime description for spatial Weyl-symmetry \cite{Gomes:2010fh}. This is shown in section \ref{sec:SDconstruction}.  The physical predictions derived form the resulting Shape Dynamics description of gravity are indistinguishable form the predictions of the standard ADM description of gravity \cite{Koslowski:2012uk}. The ontology of gravity is however changed into a theory of evolving spatial conformal geometries. This Shape Dynamics description implements Mach's principle of relativity of local scale and of the relativity of a local frame of reference in space and posses a preferred notion of simultaneity. 

The Shape Dynamics description by itself does not change the theory of gravity. Classical Shape Dynamics is locally indistinguishable form General Relativity. The situation can, to an extent, be compared to the effect of Special Relativity on Maxwell's equations. Special Relativity does not change the predictions of electromagnetism but it adds a new interpretation of the equations in terms of the spacetime picture. This opened historically the door for significant progress in theoretical physics. It is the hope that the spatial conformal geometry picture opens the door for some progress in theoretical physics as well, in particular for quantum gravity. This hope is founded in the fact that the description of gravity as a dynamical theory of spatial conformal geometry suggest new approaches to quantum gravity based on spatial conformal geometry rather than spacetime geometry. 

We start this exploration in section \ref{sec:2+1} by considering pure gravity on a torus in 2+1 dimensions, which is a completely solvable system \cite{Moncrief:1989dx,Carlip:2004ba}. We consider in particular the interpretation of causal dynamical triangulation simulations of the 2+1 torus universe. These are in complete agreement with Shape Dynamics but not with the spacetime description of General Relativity \cite{Budd:thesis}. We caution however, because pure gravity on the torus in 2+1 dimensions is very special, as it is geodesic theory on Teichm\"uller space. One should thus be careful not to generalize observations from this model to full General Relativity in 3+1 dimensions, which is vastly more complicated.  

Consequences of the Shape Dynamics description of General Relativity in 3+1 dimensions are discussed in section \ref{sec:effective}. We start with standard effective field theory, in particular with the space of effective action functionals on which exact renormalization group equations based on Shape Dynamics can be formulated. This so--called theory space is usually defined by specifying a field content, gauge symmetry principle and locality principle (for a review of the application to gravity see \cite{Niedermaier:2006wt}),. The Shape Dynamics description specifies the field content as the spatial metric and the gauge symmetries as spatial diffeomorphisms and spatial conformal transformations, rather than a theory space based on spacetime geometry. This suggests a space of effective modified gravity theories that is similar to the modified gravity theories motivated by Ho\v{r}ava-Lifshitz gravity as introduced in \cite{Horava:2009uw}. 

Also in section \ref{sec:effective}, we consider the Loop- and polymer- quantization of pure Shape Dynamics \cite{Koslowski:2013vha}. It turns out that one can construct a physical Hilbert space for the theory. This is progress compared to approaches based on the spacetime description of General Relativity. However, the complicated structure of the Shape Dynamics Hamiltonian complicates the quantization of dynamics significantly. We complete the discussion of effective Shape Dynamics by discussing how Shape Dynamics solves the problem of time and the problem of ``wrong sign'' kinetic term for the conformal mode and present an interpretation of holographic renormalization in light of Shape Dynamics following \cite{Gomes:2011dc,Gomes:2013uk}.

\section{Minimal and Non--minimal Descriptions}\label{sec:Minimal}

The main observation that underlies this contribution is the simple fact that the same classical physical system can be obtained from a number of underlaying first principles. This simple observation can have profound consequences for the quantization of the classical system, since the quantization of a classical system is often concerned with the implementation of these first principles at the quantum level. This is particularly true for the canonical Dirac quantization approach.

\subsection{Classical Systems}

The minimal content of a mathematical description of a classical physical system has to specify its set of objective\footnote{The attribute ``objective'' is not necessary at this stage, but will be important in the description of a closed universe \cite{Barbour:2013goa}.} observables $\{O_i\}_{i \in \mathcal I}$ and express the equations of motion for the observables in terms of the observables themselves\footnote{This general description is not manifestly relativistic, but it can of course be used to describe relativistic systems.}. This is to say that we require evolution equations
\begin{equation}
 \frac{\partial}{\partial t} O_i = F_i[O;t),
\end{equation}
where $t$ denotes time. A Hamiltonian system in which all phase space functions $f$ are observables fits into this definition: The set of observables is are smooth functions on phase space $\Gamma$, and the equations of motion are $\dot f=\{f,H\}$, where $H\in C^\infty(\Gamma)$ denotes the Hamiltonian of the system. 

Most descriptions of physical systems are not minimal in the sense that the system is described using auxiliary concepts. This is in particular true for gauge theories, where  auxiliary ``pure gauge'' degrees of freedom are introduced to make the description of a field theory manifestly local\footnote{I use locality in the sense that the Poisson-bracket of two observables with disjoint support vanishes and that infinitesimal time evolution preserves locality.}. The canonical description of gauge theories \cite{Henneaux:1992ig} uses a regular and irreducible set of first class constraints $\{\chi_a\}_{a\in \mathcal A}$: The observables of a gauge theory are equivalence classes of gauge--invariant phase space functions, i.e. $\{O_i,\chi_a\}=0$ for all $a\in \mathcal A$, where two representative phase space functions are considered equivalent if they coincide on the constraint surface, i.e. on the subspace of phase space where $\chi_a=0$ holds for all $a\in\mathcal A$. The equations of motion are generated by a 
total Hamiltonian 
that is the sum of a gauge--invariant $H_o$ and an arbitrary linear combination of gauge--generators $\chi_a$. 

The minimal description of a canonical gauge theory consists of a completely gauge--fixed description: For this one introduces a regular and irreducible set of gauge-fixing conditions $\{\sigma^a\}_{a \in \mathcal A}$, such that the surface defined by the vanishing of all $\sigma^a$ intersects each gauge orbit exactly once. The reduced phase space $\Gamma_o$ is the intersection of the constraint surface with the gauge--fixing surface, i.e. the subspace of $\Gamma$ where $\chi_a=0=\sigma^a$ holds for all $a\in\mathcal A$. The observables of a gauge theory are readily identified with sufficiently smooth functions on $\Gamma_o$ and the restriction $H_{gf}=H_o|_{\Gamma_o}$ generates evolution through the Dirac bracket, which is the projection of the Poisson bracket to the constraint surface. 

Given a completely gauge-fixed system, one can recover the gauge-theoretic description by parametrizing the action of the group of gauge transformations. This requires an enlargement of the phase space from $\Gamma_o$ back to $\Gamma$ and the introduction of constraints $\chi_a$, which are just the generators of the action of the gauge group. It is now very important to notice that this gauge--unfixing is not unique: Rather, there is a huge arbitrariness how one should enlarge phase space and even given the same enlarged phase space, there are in general many symmetries that can be parametrized. An example for this is free electromagnetism. Given appropriate boundary conditions, one can gauge--fix the system with axial gauge and work out the completely gauge-fixed description and subsequently gauge--unfix a shift symmetry in the electric field (see e.g. \cite{Gomes:2012hq}). The shift-symmetric description of electromagnetism exists on the phase space of the Maxwell description, but is based on a different 
set of gauge symmetries. 

The reason why a bare description is rarely used can already be seen in free electromagnetism: the bare physical degrees of freedom are non--local, while the Maxwell description is manifestly local. The locality of electromagnetism, although not manifest in the gauge--fixed description, can still be recovered from the gauge--fixed description: a local excitation will propagate within its future light--cone. 

The lesson of this section is that physical properties of the system, such as the number of degrees of freedom per point and causality should be disentangled form properties of the description of the system, such as $U(1)$--gauge invariance of electromagnetism. The physical properties of the system can be recovered in every description of the system, but may not be manifestly visible in a given description. However, properties of a non--minimal description are in general not physical properties. It is therefore possible to find different, even mutually exclusive, construction principles for the same classical system. This ambiguity can have important consequences for Dirac quantization, because one construction principle may be quantizable while another may not be quantizable.

\subsection{Classical Example}

Let us consider a well-studied bare description that can obtained through at least two distinct classical construction principles: The bare description is a Liouville field $\phi$ on a flat spatial circle. The system can be described using the charge algebra
\begin{equation}\label{equ:chargeAlgebra}
 \{Q(n_1,\xi_1)Q(n_2,\xi_2)\}=Q(N,\Xi)+K[n_1,\xi_1;n_2,\xi_2],
\end{equation}
where $Q(n,\xi)=S(n)+H(\xi)$ and
\begin{equation}
 \begin{array}{rcl}
   S(n)&=&\frac 1 2 \int d \theta\,n\,\left(\pi^2+\frac{1}{l^2}(\phi^\prime)^2+\frac{m}{c^2}e^{c\,\phi}-\frac {4}{l^2c}\phi^{\prime\prime}\right)\\
   H(\xi)&=&\int d\theta\,\xi\,\left(\pi\phi^\prime-\frac 2 c \pi^\prime\right)\\
   N&=&n_1\xi_2^\prime+\xi_1 n_2^\prime-n_2\xi_1^\prime-\xi_2 n_1^\prime\\
   \Xi&=&\xi_1\xi_2^\prime-\xi_2\xi_1^\prime+\frac 1{l^2}(n_1n_2^\prime-n_2n_1^\prime)\\
   K[n_1,\xi_1;n_2,\xi_2]&=&-\frac{4}{l^2c^2}\int d\theta\left(n_1^{\prime\prime}\xi_2^{\prime}-n_2^{\prime\prime}\xi_1^{\prime}\right),
 \end{array}
\end{equation}
where $\pi$ denotes the momentum conjugate to $\phi$ and where primes denote derivatives w.r.t. the spatial coordinate $\theta$. The dynamics of the system is encoded in the charge algebra which contains the Hamiltonian
\begin{equation}\label{equ:LiouvilleHam}
 H=Q(1,0)=\frac 1 2 \int d\theta\left(\pi^2+\frac{1}{l^2}(\phi^\prime)^2+\frac{m}{c^2}e^{c\,\phi}\right),
\end{equation}
where $\pi$ denotes the momentum conjugate to $\phi$ and primes denote derivatives w.r.t. the spatial coordinate $\theta$. Using the mode expansion
\begin{equation}
 L^\pm_{\pm m}:=Q\left(\frac{i\pm 1}{2}e^{im\theta},l\frac{i\pm 1}{2}e^{im\theta}\right)
\end{equation}
one finds that the Poisson-brackets satisfy a pair of Virasoro algebras
\begin{equation}\label{equ:VirasoroAlg}
 \begin{array}{rcl}
   i\{L^\pm_{\pm m},L^\mp_{\mp n}\}&=&0\\
   i\{L^\pm_{\pm m},L^\pm_{\pm n}\}&=&(m-n)L^\pm_{m+n}+\frac{4\pi}{c^2 l}m^3 \delta_{m,-n},
 \end{array}
\end{equation}
from which we read-off value of the central charges as $C=\frac{48 \pi}{c^2 l}$. I will now, for the sake of clarity, abuse notation and call the classical charge algebra (\ref{equ:VirasoroAlg}) with Hamiltonian (\ref{equ:LiouvilleHam}) Liouville theory. This is not entirely correct as the elementary degrees of freedom are almost, but not completely encoded in (\ref{equ:VirasoroAlg}). 

\subsubsection{Spacetime Description}

A particularly interesting way to construct the bare system, which is encoded in the charge algebra (\ref{equ:VirasoroAlg}) was found by Brown and Henneaux \cite{brown-henneaux} and can be viewed as a precursor of the AdS/CFT duality. It is based on General Relativity in 2+1 spacetime dimensions with asymptotically AdS boundary conditions. The fundamental field is the spacetime metric $g_{\mu\nu}$ and the symmetries of the theory are spacetime diffeomorphisms. It follows form the fact that diffeomorphisms are gauge symmetries that there are no local gravitational degrees of freedom, so all degrees of freedom on a topologically trivial domain are boundary degrees of freedom, which we will now consider. 

Using the time variable $t$ and polar coordinates $(r,\phi)$, the asymptotically AdS boundary conditions of \cite{brown-henneaux} are:
\begin{equation}
 \begin{array}{lll}
   g_{tt}=-\frac{r^2}{l^2}+O(1),&g_{tr}=O(r^{-3}),&g_{t\phi}=O(1)\\
   g_{rr}=\frac{l^2}{r^2}+O(r^{-4}),&g_{\phi\phi}=r^2+O(1),&g_{r\phi}=O(r^{-3}).
 \end{array}
\end{equation}
The asymptotic space--time symmetries, i.e. spacetime--diffeomorphisms that preserve the boundary conditions modulus those that vanish at the boundary, are parametrized by functions on a circle $f^\pm=f(\phi\pm\frac{t}{l})$; the explicit form of the generating vector fields $u[f^\pm]$ is
\begin{equation}
  u^t[f^\pm]=\frac{l}{2}\left(f^\pm+\frac{l^2}{2r^2}f^\pm_{,\phi\phi}\right),\,\,\,\,u^r[f^\pm]=\mp\frac r 2 f^\pm_{,\phi},\,\,\,\,u^\phi[f^\pm]=\pm\frac 1 2\left(f^\pm-\frac{l^2}{2r^2}f^\pm_{,\phi\phi}\right).
\end{equation}
These vector--fields are defined only up to terms that vanish in the limit $r\to\infty$ and satisfy the following Lie--algebra
\begin{equation}\label{equ:LieAlg}
   [u[f_1^\pm],u[f_2^\mp]]=0,\,\,\,\,\,\,\,\,\,\,
   [u[f_1^\pm],u[f_2^\pm]]=u[f_1^\pm f^\pm_{2,\phi}-f^\pm_2f^\pm_{1,\phi}].
\end{equation}
A consequence of the fact that the gauge--generators of of spacetime diffeomorphisms $D(u)$ acquire explicit boundary terms $B(u)$ is that the algebra of diffeomorphism generators is a central extension of the Lie--algebra (\ref{equ:LieAlg}), i.e. it takes the form $\{D(u_1),D(u_2)\}=D([u_1,u_2])+K(u_1,u_2)$. There is a large ambiguity in the precise choice of boundary term, which can be used to make all boundary terms vanish for one particular reference spacetime, which we can choose to be pure AdS space. An expansion of $f^\pm$ in Fourier modes then gives a pair of Virasoro algebra with central charge
\begin{equation}
  C=\frac{3l}{2G}.
\end{equation}
Since the entire physical content of asymptotically AdS gravity is contained in this pair of Virasoro algebras, we can identify the bare system, i.e. Liouville theory, with the asymptotically AdS gravity, whenever the central charges match. We have thus found a non--minimal description of Liouville theory which tells us an underlaying construction principle for Liouville theory as Einstein spacetimes which satisfy asymptotically AdS boundary conditions. To examplify that the non-minimal descriptions are not unique, we will now consider a different non-minimal description of Liouville theory.

\subsubsection{Spatially conformal Description}

We will now consider a dynamical theory of a spatial metric $g_{ab}$ in 2 dimensions that implements spatial diffeomorphism- and spatial Weyl- invariance through the constraints:
\begin{equation}
 \begin{array}{rcl}
   H(v)&=&\int_\Sigma d^2x \,\pi^{ab}(\mathcal L_v g)_{ab}=\int_{\partial \Sigma} dx_c v^c g_{ab}\pi^{ab}-\int_\Sigma d^2x\, g_{ab}(\mathcal L_v \pi)^{ab}\\
   C(\rho)&=&\int_\Sigma d^2x\,\rho\, g_{ab}\pi^{ab}, 
 \end{array}
\end{equation}
where $\pi^{ab}$ denotes the canonically conjugate momentum density. These constraints imply that all local degrees of freedom are pure gauge, since any spatial metric on a topologically trivial domain is diffeomorphic to a conformally flat metric. All physical degrees of freedom of the system are thus boundary degrees of freedom. Another consequence is that the Hamiltonian over a topologically trivial domain depends only on the boundary values of the fields. We will choose an appropriate Hamiltonian shortly.

Using polar coordinates $(r,\phi)$ on $\Sigma$ we impose canonical fall--off conditions that are satisfied in $t=const$ slices of the asymptotically AdS spacetimes we considered in the previous section:
\begin{equation}\label{equ:fall-off}
 \begin{array}{ccc}
   g_{rr}=\frac{l^2}{r^2}+O(1),&g_{r\phi}=O(r^{-3}),&g_{\phi\phi}=r^2+O(1),\\
   \pi^{rr}=O(r^{-1}),&\pi^{r\phi}=O(r^{-2}),&\pi^{\phi\phi}=O(r^{-5}).
 \end{array}
\end{equation}
The asymptotic symmetries, i.e. the conformal diffeomorphisms that leave the boundary conditions invariant modulus those that vanish at the boundary, can be parametrized by a function $\alpha(\phi)$:
\begin{equation}\label{equ:ASG}
 u^r[\alpha]=-r\alpha^\prime+O(r^{-1}),\,\,\,\,\,\,\,\,u^\phi[\alpha]=\alpha-\frac{l^2}{2r^2}\alpha^{\prime\prime}+O(r^{-4}),
\end{equation}
where primes denote derivatives w.r.t. $\phi$ and where the lower orders vanish in the boundary limit $r\to\infty$. The vector fields satisfy the Lie--algebra
\begin{equation}\label{equ:spatialLieAlg}
 [u[\alpha],u[\beta]]=u[\alpha\beta^\prime-\beta\alpha^\prime]
\end{equation}
in the limit $r\to\infty$. The conformal constraints $C(\rho)$ are ultralocal and do not contribute to the asymptotic symmetries, but parts of the constraints $H(v)$ generate asymptotic symmetries require boundary terms; a sufficient boundary for the vector fields (\ref{equ:ASG}) is $B(u)=2\int_{\partial \Sigma} dx_c u^ag_{ab}\pi^{bc}$, where the momentum densities are pulled-back to $\partial \Sigma$. Using that the variation of $H(u)+B(u)$ reduces by construction to the bulk-variation of $H(u)$, we find that
\begin{equation}
 \begin{array}{l}
   \{H(u)+B(u),H(v)+B(v)\}-H([u,v])\\=\int_{\partial \Sigma} dx_c\left(\pi^{ab}(v^c\mathcal L_u-u^c\mathcal L_v)g_{ab}+2[u,v]^ag_{ab}\pi^{bc}\right).
 \end{array}
\end{equation}
Using that the  vector fields (\ref{equ:ASG}) are asymptotic symmetries, we find in the boundary limit $r\to\infty$ that $\int_{\partial \Sigma} dx_c\,\pi^{ab}(v^c\mathcal L_u-u^c\mathcal L_v)g_{ab}$ vanishes. We thus have $\{H(u)+B(u),H(v)+B(v)\}=H([u,v])+B([u,v])$ when $u,v$ are vector fields of the form (\ref{equ:ASG}), so the algebra for the asymptotic symmetries is not deformed by the boundary terms. We thus find the physical ``momentum'' variables $J[\alpha]:=H(u[\alpha])+B(u[\alpha])$ with algebra
\begin{equation}
 \{J[\alpha],J[\beta]\}=J[\alpha\beta^\prime-\beta\alpha^\prime].
\end{equation}
The canonically conjugate ``configuration'' variables are asymptotic values of the gauge-fixing conditions $G[\beta]$, which, after ignoring the global mode, we can choose as
\begin{equation}
 G[\beta]=K+\int_{\partial \Sigma} dx_c \bar{G}^{abc}\left(\beta g_{ab\bar{;}c}-\beta_{,c}(g_{ab}-\bar{g}_{ab})\right),
\end{equation}
where $G$ denotes DeWitt's supermetric and bars denote the pure AdS metric and where $K[\beta]$ denotes the boundary limit of the pure AdS Hamiltonian plus counter term smeared against $\beta$. The Dirac algebra of  was also derived in \cite{brown-henneaux} and it turns out to yield a pair of Virasoro algebras with central charge $C=\frac{3l}{2G}$.

We have thus found an independent construction principle for Liouville theory which is not based on spacetime, provided we choose the coupling constant such that the central charges match. Rather, the second construction principle is based on a dynamical theory of the spatial metric, which posses spatial diffeomorphism- and spatial Weyl-invariance and canonical fall--off conditions (\ref{equ:fall-off}). This shows that the auxiliary concepts, which are on the one hand spacetime geometry and spacetime diffeomorphism invariance and on the other hand dynamical spatial geometry and spatial conformal diffeomorphism invariance, can be interchanged without changing the bare theory, which is Liouville theory on a circle.

\subsection{Quantum Systems}

We saw how the same bare classical system can be constructed form very distinct construction principles. This means that the physical interpretation of one and the same bare classical can be very different. A similar ``freedom of interpretation'' exists in quantum systems. 

To explain this freedom, we define a quantum system as a dynamical system $(\mathfrak A,U,\mathbb G)$, consisting of a normed $*$-algebra $\mathfrak A$ and a unitary action $U$ of a group $\mathbb G$ as automorphisms of $\mathfrak A$, where a one-dimensional subgroup $G$ is interpreted as generating time--evolution through $U_t$. In the simplest case we have a Hilbert space $\mathcal H$ in the background and $\mathbb G=\mathbb R$ and $U_t:a \mapsto e^{-iHt}\,a\,e^{iHt}$, for some Hamiltonian $H$ on $\mathcal H$. Such a bare quantum system does not admit a classical limit. To be able to define a classical limit, one needs at least a notion of ``configuration operators,'' i.e. a algebra morphism $i$ from a commutative algebra $C(\mathbb X)$, where $\mathbb X$ denotes a configuration space, into a commutative subalgebra $\mathfrak A_o$ of $\mathfrak A$, which provides an interpretation of the elements of $\mathfrak A_o$ as configuration operators. Given a state, i.e. a continuous positive linear functional $\
omega$, on $\mathfrak A$, we can derive the time evolution of all configuration operators $a_o\in\mathfrak A_o$ in this state as $\omega(U_t(a_o))$, which may allow us to derive a semiclassical interpretation if $\omega$ is a semiclassical state for $\mathfrak A_o$.

However, in general, there is a huge amount of ambiguity in constructing a tuple $(\mathbb X, i:C(\mathbb X)\to \mathfrak A)$, even if one require that the image of $i$ is a maximal commuting subalgebra of $\mathfrak A$. It follows that the same bare quantum system $(\mathfrak A,U,\mathbb G)$ can have many distinct classical limits, each coming form a different physical interpretation of the elements of the operator algebra $\mathfrak A$. We will now give a simple example showing the ambiguity of the classical limit of a bare quantum system.

\subsection{Quantum Example}

A very simple example that shows how the same bare quantum system is interpreted as two distinct classical systems by identifying different sets of ``elementary'' operators is obtained by the bare description of the harmonic oscillator. The bare system is most easily constructed on the Hilbert space $\mathcal H=l^2$, the space of square summable complex sequences $(a_n)_{n=0}^\infty$ where $\sum_{n}|a_n|^2<\infty$. The operator algebra is given by bounded linear operators $B(l^2)$, but it is closer to standard physical notation, if we work with unbounded densely defined generators. For this we consider a dense basis $\{e_n\}_{n=0}^\infty$ and define the unbounded generators as linear operations of the basis elements. We choose the transformation group $\mathbb G=\mathbb R$, which is physically interpreted as time evolution. The generator of time evolution is defined through the map: $H: e_n \mapsto \omega_o (n+\frac 1 2) e_n$, so the time evolution is
\begin{equation}
 U_t: e_n \mapsto e^{i\omega_o (n+\frac 1 2)t} e_n. 
\end{equation}
This defines the bare quantum system.

This bare system is readily interpreted as the harmonic oscillator, if we impose the physical interpretation of the linear maps
\begin{equation}
 a:e_n\mapsto \sqrt{n+1}\,e_{n+1},\,\,\,\,\,\,a^*:e_n\mapsto \sqrt{n}e_{n-1}
\end{equation}
as the standard one--dimensional ladder operators. Using this physical interpretation, the Hamiltonian $H$ takes the form
\begin{equation}
 H=\epsilon(a^*a+\frac 1 2),
\end{equation}
which is the standard representation of the harmonic oscillator.

A different physical interpretation of the same Hamiltonian is most easily described using the bijection $(j,m):=e_{j^2+j+m}$ where $j\in \mathbb N_o$ and $m=-j,...,j$. Using this relabeling of the basis, the second physical interpretation is obtained through the interpretation of the linear maps
\begin{equation}
 \begin{array}{rccl}
  J_+:&(j,m) &\mapsto& \sqrt{(j-m)(j+m+1)}\,(j,m+1)\\
  J_-:&(j,m) &\mapsto& \sqrt{(j+m)(j-m+1)}\,(j,m-1),\\
  J_3:&(j,m) &\mapsto& m (j,m)
 \end{array}
\end{equation}
as the standard three--dimensional angular momentum operators $J_i$. The Hamiltonian $H$ then takes the form
\begin{equation}
 H=\epsilon(\vec{J}^2+J_3+\frac 1 2),
\end{equation}
which is recognized as a particle on a 2-sphere in an external magnetic field. This example shows how the same bare quantum system can be interpreted in a number of distinct ways simply through reinterpreting the roles of so-called ``elementary'' operators. 

We can now reverse the logic of this example and view the bare quantum system as a recognizable or ``standard'' quantization of two distinct classical systems, which do not appear to be related at first sight. In particular, knowing the relation between the harmonic oscillator and the particle on the sphere, we can quantize one system and reinterpret it as the quantization of the other. This leads to the main message of this section: the minimal description of a classical or a quantum system can be obtained from many different first principles. So, if one set of first principles can not be implemented, one should look for a different set which can be implemented. This is the approach that we will subsequently propose for quantum gravity, where it appears that the spacetime principles of General Relativity can not be implemented within a standard quantum field theory context. Rather than abandoning the bare gravitational system, we will propose to base the quantization of gravity on a set of principles that 
is founded in a different set of construction principles based on the Shape Dynamics description of gravity. In the next section, we will outline how the construction principle for shape dynamics differs form the construction principle of General Relativity.

\section{Spatial conformal and Spacetime description of gravity}\label{sec:Ontology}

Before presenting the construction of Shape Dynamics in the next section, I will contrast the spacetime picture that underlies the construction of General Relativity with a conceptual picture based on spatial Weyl--invariance, which we will use to derive Shape Dynamics. A remarkable result by Hojman, Kucha\v{r} and Teitelboim \cite{Hojman:1976vp} shows that the canonical formulation of General Relativity can be derived entirely form spacetime considerations. This construction starts with the consideration of the Lie--algebra of spacetime vector fields and the pull--back of this Lie--algebra to a 3-dimensional Riemannian hypersurface. The 3+1--decomposition into a shift part $v$ and a lapse part $N$ provides Dirac's hypersurface--deformation algebra
\begin{equation}
  [H(v_1),H(v_2)]=H([v_1,v_2]),\,\,\,\,[H(v),S(N)]=S(\mathcal L_v N),
\end{equation}
\begin{equation}
  [S(N_1),S(N_2)]=H(N_1\nabla N_2-N_2\nabla N_1),
\end{equation}
where $H(.)$ and $S(.)$ denote canonical generators built from the spatial metric $g_{ab}$ and its canonically conjugate momentum density $\pi^{ab}$. The form of the generators $H(v)$ is determined by $[H(v_1),H(v_2)]=H([v_1,v_2])$ and the requirement that these generators generate infinitesimal spatial diffeomorphisms $\delta_v g_{ab}=(\mathcal L_v g)_{ab}$, $\delta_v \pi^{ab}=(\mathcal L_v \pi)^{ab}$ to be
\begin{equation}
 H(v)=\int d^3x\, \pi^{ab}(\mathcal L_v g)_{ab}.
\end{equation}
Then, assuming time--reversability and a strong notion of locality \cite{Hojman:1976vp} proves a representation theorem for the generators $S(N)$, which are shown to be of the form
\begin{equation}
 S(N)=\int d^3x N\left(\frac{\pi^{ab}G_{abcd}\pi^{cd}}{\sqrt{|g|}}-(R-2\Lambda)\sqrt{|g|}\right),
\end{equation}
where $G_{abcd}=g_{ac}g_{bd}-\frac 1 2 g_{ab}g_{cd}$ denotes the DeWitt supermetric , $R$ the scalar curvature of $g_{ab}$, $|g|$ the determinant of $g_{ab}$ and where $\Lambda$ is an undetermined constant. This is precisely the form of the ADM constraints, as they are obtained from the Legendre--transform of the Einstein-Hilbert action, which encode the dynamics of the canonical description of General Relativity. This is why this result is interpreted as a derivation of the dynamics of gravity from space--time first principles.

A different set of first principles to derive the dynamics of canonical gravity is based on conceptual ideas by Barbour and collaborators\cite{Barbour:1982gha,Barbour:2011dn,Anderson:2004bq,Anderson:2004wr,Barbour:2003gr,Barbour:2000qg}. These authors who propose that gravity is be a dynamical theory of spatial conformal geometry and a preferred notion of simultaneity. The conceptual postulates can be summarized as follows:
In a closed system, such as the entire universe, one should not need external clocks and rods to interpret physics, as used in the interpretation of spacetimes. Rather, clocks and rods should be replaced with local dimensionless ratios of physical quantities. To implement this conceptual idea, Barbour and collaborators used two construction principles for classical theories: 
\begin{enumerate}
 \item {\it Manifest reparametrization invariance} can be implemented into the dynamics of a particle system with energy $E$, kinetic energy $T=\dot x^a h_{ab}(x)\dot x^b$ and potential $V(x)$ by replacing $dt \to N(\lambda) d\lambda$ and treating $N$ as a Lagrange--multiplier. The elimination of $N$ through its equations of motion yields
 \begin{equation}
   S=\int d\lambda \sqrt{\left(E-V(x)\right)\,\frac{\partial x^a}{\partial \lambda}h_{ab}(x)\frac{\partial x^b}{\partial \lambda}},
 \end{equation}
 which is manifestly invariant under affine reparametrizations $\lambda \to f(\lambda)$ and hence removes all dependence of predictions of the system from the choice of time--parametrization. The square root in the reparametrizaton--invariant form arises because the kinetic term $T$ was assumed to be quadratic in velocities and allows us to interpret the action $S$ as a geodesic length on configurations space. 
 \item {\it Best matching} is based in the idea that some transformations $T:x^a\mapsto T^a_b(\phi)x^b$, where $\phi^\alpha$ denotes a parametrization of the transformation group, should not alter the physical state, i.e. these transformations should be pure gauge. One can implement this by extending configuration space with the  the transformation group (i.e. the extended configuration space is the product of the original configuration space and the group manifold) and replacing $x^a \to T^a_b(\phi) x^b$ everywhere in the action. This procedure changes the velocities to $\frac{\partial x^a}{\partial \lambda} \to T^a_b(\phi)\frac{\partial x^b}{\partial \lambda}+{T^a_b}_{,\alpha} x^b \frac{\partial \phi^\alpha}{\partial \lambda}$. The equations of motion for the group parameters $\phi^\alpha$ then implement the desired invariance.
\end{enumerate}
These two construction principles can now be used to construct actions for theories of the spatial metric that (a) dynamical theories of the spatial conformal geometry, i.e. spatial diffeomorphisms and spatial Weyl--transformations are pure gauge and (b) invariant under local reparametrizations of time. If we assume a kinetic term $T(x)$ that is quadratic in velocities, then local time--reparametrization invariance is implemented by writing the action as
\begin{equation}
 S=\int d\lambda d^3x \sqrt{V(x)T(x)}.
\end{equation}
This form of the action ensures local time--reparametrization invariance and thus implements the physical idea that clocks should be built from objective local observables. For definitiveness, we choose a potential $V=(R-2\Lambda)\sqrt{|g|}$. It turns out to be straightforward to implement best--matching w.r.t. spatial diffeomorphisms, which turns the theory into a dynamical theory of the spatial geometry. 

Best matching w.r.t. spatial conformal transformations turns out to be more subtle for two reasons: 

(1) It turns out that equivalence with the dynamics of General Relativity on a closed manifold requires one to best match w.r.t. conformal transformations that do not change the total spatial volume, i.e. we restrict our derivation to conformal factors $e^{\hat \phi(x)}=\hat \Omega(x)=\langle\Omega^6\rangle^{-\frac{1}{6}}\Omega(x)$, where $\langle f \rangle:=\frac{\int d^3x \sqrt{|g|} f}{\int d^3y \sqrt{|g|}}$. The Legendre--transform of the best--matched theory yields the constraints
\begin{equation}\label{equ:LTConstraints}
 \begin{array}{l}
   S(N)=\int d^3x\,N\left(\frac{\sigma^a_b\sigma^b_a}{\sqrt{|g|}}\hat{\Omega}^{-6}+(2\Lambda-\frac 1 6 \langle \pi\rangle^2)\sqrt{|g|}\hat \Omega^6-\hat \Omega(R-8\Delta)\hat \Omega\sqrt{|g|}\right)\\
   H(v)=\int d^3x\,\left(\pi^{ab}(\mathcal L_v g)_{ab}+\pi_\phi \mathcal L_v \phi\right)\\
   C(\rho)=\int d^3x\,\rho\left(\pi_\phi-4\left(\pi-\langle\pi\rangle\sqrt{|g|}\right)\right)\\
   Q(\sigma)=\int d^3x \,\sigma\,\pi_\phi,
 \end{array}
\end{equation}
where $\sigma^a_b$ denotes the trace--free part of $g_{ac}\pi^{cb}$. We see from the first line of (\ref{equ:LTConstraints}), that best--matching w.r.t. the restricted set of conformal transformations has the profound consequence that all terms in the $S(N)$ obtain York--scaling. This is very important, because this means that one can use these first principles to derive the constant mean (extrinsic) curvature (CMC) gauge--fixed version of canonical gravity. 

(2) Even after the restriction to local conformal transformations that do not change the total spatial volume, one sees that the best matched theory (\ref{equ:LTConstraints}) is not a gauge theory under this restricted set of conformal transformations. This can be seen from the fact that the constraints $S(N)$ together with the $Q(\sigma)$ are a second class system. In the next two sections we will remedy this fact using the symmetry trading mechanism.

\section{Symmetry Trading and Symmetry Doubling}\label{sec:symmetryTrading}

We have already seen that gauge symmetries are a property of the description of a physical system rather than a physical property. I will now explain the symmetry trading \cite{Gomes:2011zi} and symmetry doubling mechanisms \cite{Gomes:2012hh}, which we will subsequently use to construct a description of gravity as a dynamical theory of the spatial conformal geometry.

\subsection{Symmetry Trading}

Let us consider a classical gauge theory, i.e. the data $(\Gamma,\{.,.\},H,\Xi)$ consisting of a phase space $\Gamma$ with Poisson bracket $\{.,.\}$, Hamiltonian $H$ and a set of first class constraints $\Xi=\{\chi^\mu\}_{\alpha\in\mathcal M}$, which we assume to be regular and irreducible. The first class $\chi^\mu$ are the generators of gauge transformations. We will now proceed with the construction of what is called a special linking theory in \cite{Gomes:2011zi}. For this we assume that the phase space $\Gamma$ can be written as a product of two Poisson--commuting subspaces $\Gamma_o$ and $\Gamma_e$. For concreteness, we will introduce local coordinates in the form of canonical pairs $(q^a,p_a)$ on $\Gamma_o$ and $(\phi^\alpha,\pi_\alpha)$ on $\Gamma_e$; the non--vanishing elementary Poisson--bracket on $\Gamma_o$ is $\{q^a,p_b\}_o=\delta^a_b$ and the non--vanishing elementary Poisson--bracket on $\Gamma_e$ is $\{\phi^\alpha,\pi_\beta\}_e=\delta^\alpha_\beta$. A special linking theory is now a theory 
where the constraint surface $\mathcal C=\{x \in \Gamma: \chi^\mu(x)=0\,\forall \mu \in \mathcal M\}$ can specified through
\begin{equation}
  \begin{array}{rcl}
    0&=&\phi^\alpha-\chi_1^\alpha(q,p)\\
    0&=&\pi_\alpha-\chi^2_\alpha(q,p)\\
    0&=&\chi_3^\nu(q,p),
  \end{array}
\end{equation}
which is to say that the constraints $\Xi$ are equivalent to three sets of constraints, where the first is equivalent to $\Xi_1=\{\phi^\alpha-\chi_1^\alpha(q,p)\}_{\alpha \in \mathcal A}$, the second is equivalent to $\Xi^2=\{\pi_\alpha-\chi^2_\alpha(q,p)\}_{\alpha \in \mathcal A}$ and and the third is equivalent to $\Xi_3=\{\chi_3^\nu(q,p)\}_{\nu \in \mathcal N}$. We thus require that the first set can be uniquely solved for the $\phi^\alpha$ and that the second can be uniquely solved for the $\pi_\alpha$ and that the third is weakly independent of $\Gamma_e$. In this situation, there are two preferred sets of partial gauge--fixing conditions:
\begin{equation}
 \Sigma^1=\{\pi_\alpha\}_{\alpha \in \mathcal A}\,\,\,\textrm { and }\,\,\,\Sigma_2=\{\phi^\alpha\}_{\alpha \in \mathcal A}.
\end{equation}
The phase space reductions associated with these two gauge fixing conditions are
\begin{equation}
  \begin{array}{rcl}
    \Sigma^1&:& (\phi^\alpha,\pi_\beta)\,\to\,(\chi_1^\alpha,0)\\
    \Sigma_2&:& (\phi^\alpha,\pi_\beta)\,\to\,(0,\chi^2_\beta),
  \end{array}
\end{equation}
so the reduced phase space for either gauge--fixing is $\Gamma_o$. The Dirac bracket associated with either of these phase space reductions coincides with the Poisson--bracket
\begin{equation}
 \{.,.\}^1_D=\{.,.\}_o\,\,\,\textrm{ and }\,\,\,\{.,.\}^2_D=\{.,.\}_o,
\end{equation}
showing that both reductions lead to $(\Gamma_o,\{.,.\}_o)$. However, the Hamiltonian and the reduced constraint do not coincide, rather the first phase space reduction leads to
\begin{equation}
 \begin{array}{rcl}
   H^1&=&H|_{\phi^\alpha=\chi_1^\alpha,\pi_\beta=0}\\
   \tilde \Xi_1&=&\{\chi_1^\alpha\}_{\alpha \in \mathcal A} \cup \Xi_3
 \end{array}
\end{equation}
while the second phase space reduction leads to
\begin{equation}
 \begin{array}{rcl}
    H^2&=&H|_{\phi^\alpha=0,\pi_\beta=\chi^2_\beta}\\
   \tilde \Xi_2&=&\{\chi^2_\alpha\}_{\alpha \in \mathcal A} \cup \Xi_3.
 \end{array}
\end{equation}
\begin{figure}
  \begin{center}
  \includegraphics[width=\textwidth]{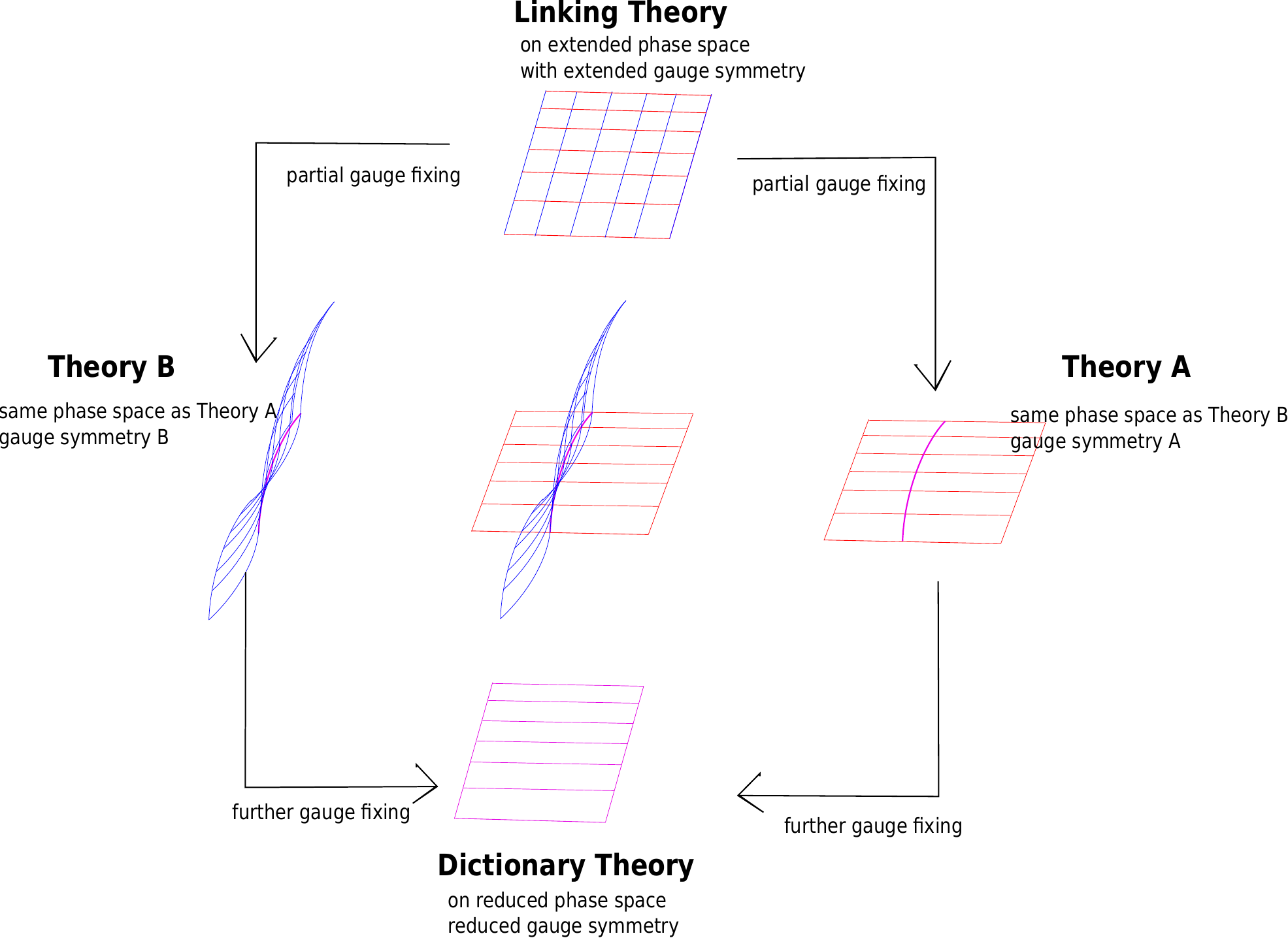}
  \end{center}
  \caption{\label{figure:LT} Two gauge theories with distinct gauge symmetries can be constructed from a linking theory through partial gauge--fixing. The bare description, i.e. the ``dictionary theory'' (purple), is the (completely) gauge--fixed linking theory. This bare description can be obtained as the intersection of two first class constraint surfaces, i.e. the red and blue surfaces which represent the two distinct gauge theories. The red and blue constraint surfaces are each first class, but have different gauge symmetries. This is illustrated by the gauge--orbits on the two constraints surfaces, which are indicated as the red and blue lines.}
\end{figure}
We thus have the same phase space $\Gamma_o$ with the same Poisson--bracket $\{.,.\}_o$, which carries two distinct gauge theories, $\mathcal G^1=(\Gamma_o,\{.,.\}_o,H^1,\tilde \Xi_1)$ and $\mathcal G^2=(\Gamma_o,\{.,.\}_o,H^2,\tilde \Xi_2)$. The two theories $\mathcal G^1,\mathcal G^2$ describe the same physics, because they are obtained as two distinct partial gauge--fixings of a single original gauge theory, i.e. the linking theory. The manifest equivalence of the two gauge theories can be seen by observing that the sets $\{\chi_1^\alpha\}_{\alpha \in \mathcal A}$ and $\{\chi^2_\alpha\}_{\alpha \in \mathcal A}$ gauge--fix each other. We can thus impose the gauge fixing conditions $\tilde\Sigma^1=\{\chi^2_\alpha\}_{\alpha \in \mathcal A}$ on $\mathcal G^1$ and $\tilde\Sigma^2=\{\chi_1^\alpha\}_{\alpha \in \mathcal A}$ on $\mathcal G^2$. These two further gauge fixings turn out to be equivalent to imposing $\Sigma^1\cup\Sigma_2$ on the linking theory. This gauge--fixed theory is a bare description of the 
physical system and provides a ``dictionary'' between the two descriptions.

The important observation is that the gauge--symmetries of $\mathcal G^1$ and $\mathcal G^2$, i.e. the finite transformations generated by the action of $\tilde \Xi_1$ and $\tilde \Xi_2$, are different despite the fact that the two theories describe the same classical evolution of the same classical observables. This equivalence is due to the fact that the bare description of the two systems is the completely gauge--fixed linking theory and that this bare description can be gauge--unfixed in many different ways. In the particular case that can be constructed from the linking theory one obtains two gauge theories whose constraint surfaces gauge--fix each other (see figure \ref{figure:LT}). It turns out that one can reverse the construction and start with two first class constraint surfaces that gauge--fix each other to construct a linking theory. One can thus trade gauge--generators for gauge--fixing conditions, if the gauge--fixing conditions are first class.

We refer to the gauge--fixed description of the system as ``dictionary'' because it provides a way to relate the observables of the two theories. To see this relation, we remember that observables of a gauge theory are gauge--invariant functions (i.e. constants along gauge orbits) on the constraint surface of a gauge theory. An observable is thus completely determined by its value on a gauge--fixing surface. An observable of $\mathcal G^1$ can thus be related to the equivalent an observable of $\mathcal G^2$ by restricting it to the gauge--fixing surface, where it coincides with precisely one observable of $\mathcal G^2$. This relation can also be worked out through the linking theory: Given an observable in the linking theory, which will be represented by an equivalence class of functions $[O]$ on $\Gamma_o\times \Gamma_e$, we find the relation
\begin{equation}
 [O_1]=[O|_{\phi^\alpha=\chi_1^\alpha,\pi_\beta=0}] \,\leftrightarrow\,[O_2]=[O_{\phi^\alpha=0,\pi_\beta=\chi^2_\beta}],
\end{equation}
where $[O_1]$ denotes an observable of $\mathcal G^1$ and $[O_2]$ denotes an observable of $\mathcal G^2$.

\subsection{Excursion: Symmetry Doubling}

The linking theory formalism allows us to trade gauge--generators for gauge--fixing conditions that are first class. It is therefore natural to consider the canonical BRST description of the system, which treats gauge--transformations and gauge--fixing conditions on the same footing \cite{Gomes:2012hh}. To keep the presentation simple, I will only cover the case of Abelian case here. However, locally in phase space, one can extend the construction to the non--Abelian case as we will see.

Given a regular, irreducible set of Abelian set of constraints $\{\chi_\alpha\}_{\alpha \in \mathcal A}$, i.e. $\{\chi^\alpha,\chi^\beta\}=0$ for all $\alpha,\beta \in \mathcal A$, we require an Abelian set of gauge--fixing conditions $\{\sigma^\alpha\}_{\alpha \in \mathcal A}$. The Hamiltonian $H_o$ of a gauge theory must be gauge--invariant, but it will in general not Poisson commute with the gauge--fixing conditions. An important special case for which this happens is a Hamiltonian that vanishes on the constraint surface $H_o|_{\chi\equiv 0}=0$. Having this case in mind, we require from now on that we find a Hamiltonian that strongly Poisson--commutes with constraints and the gauge--fixing conditions.

The BRST formalism requires to extend phase space with a ghost anti--ghost pair $(\eta_\alpha,P^\alpha)$ for each gauge generator $\chi^\alpha$, which has the opposite statistics of the constraint $\chi^\alpha$ and the construction of a BRST generator
\begin{equation}
 \Omega=\eta^\alpha\chi_\alpha+\mathcal O(\eta^2),
\end{equation}
where the $\mathcal O(\eta^2)$ is chosen such that $\{\Omega,\Omega\}=0$, which is satisfied by vanishing $\mathcal O(\eta^2)$ for Abelian constraints, but can be satisfied for any set of first class constraints. The ghost number, i.e. the number of ghosts minus the number of antighosts, defines a grading of the algebra of phase space functions. The BRST generator has ghost number $+1$ and the fact that $\{\Omega,\Omega\}=0$ allows us to define the BRST differential $s$ as 
\begin{equation}
 s: f \mapsto \{\Omega,f\},
\end{equation}
which turns the algebra of phase space functions into a graded differential algebra. The important observation that underlies the BRST--formalism is the fact that this graded differential algebra provides a resolution of the observable algebra, i.e. the observables are identified with the cohomology of the differential $s$ at ghost number zero. Time evolution is generated by the BRST gauge--fixed Hamiltonian
\begin{equation}
 H_\Psi=H_o+\eta^\alpha V^\beta_\alpha P_\beta+\{\Omega,\Psi\},
\end{equation}
where $\Psi$ is phase space function of ghost number $-1$ and where $\{H_o,\chi_\alpha\}=V^\beta_\alpha\chi_\beta$, so the $V^\beta_\alpha$ vanish if the on--shell Hamiltonian $H_o$ vanishes; in this case, we have
\begin{equation}\label{equ:BRSTHamiltonian}
 H_\Psi=s\Psi=\{\Omega,\Psi\},
\end{equation}
which is formally symmetric in the $\Omega$ and $\Psi$ and hence formally symmetric in the BRST description of gauge generators (encoded in $\Omega$) and the gauge--fixing conditions (encoded in $\Psi$). 

This symmetry between gauge--generators and gauge--fixing conditions can be combined with the observation that symmetry trading requires two first class constraint surfaces that gauge--fix each other. This means that the gauge--fixing conditions $\{\sigma^\alpha\}_{\alpha \in \mathcal A}$ are themselves a set of first class constraints. We can thus construct a ghost--number $-1$ generator $\Psi=\sigma^\alpha P_\alpha +\mathcal O(P^2)$, where the $\mathcal O(P^2)$ can be chosen such that $\{\Psi,\Psi\}=0$. This allows us to define a second differential
\begin{equation}
 \tilde s: f \mapsto \{f,\Psi\}.
\end{equation}
This differential does again provide a resolution of the observable algebra, but this time based on the gauge--transformations generated by $\{\sigma^\alpha\}_{\alpha \in \mathcal A}$. These two observable algebras do however describe the same physics according to the argument of the previous section.

We are thus able to view the BRST gauge--fixed Hamiltonian (\ref{equ:BRSTHamiltonian}) in two ways: either as $s\Psi$ or $\tilde s \Omega$. The first view suggests a gauge symmetry generated by the constraints $\chi_\alpha$, while the second view suggests a gauge symmetry generated by the $\sigma^\alpha$. We are thus in a situation in which the BRST gauge--fixed equations of motion can be interpreted as based on either set of gauge symmetries. The BRST--formalism thus ``doubles symmetry'' in the particular  sense described here. 

\section{Construction of Shape Dynamics}\label{sec:SDconstruction}

We will now apply the symmetry trading mechanism to the canonical formulation of General Relativity, the ADM formalism, to obtain a the Shape Dynamics description of gravity, which is a gauge theory of spatial diffeomorphisms and spatial conformal transformations.

\subsection{ADM in CMC gauge}

The construction of Shape Dynamics can be viewed as a gauge--unfixing of the ADM formalism in constant mean (extrinsic) curvature (CMC) gauge, which provides the dictionary between the ADM formulation and Shape Dynamics. We will therefore start with this description of GR. The total ADM Hamiltonian is a linear combination of the first class constraints:
\begin{equation}
 \begin{array}{rcl}
   S(N)&=&\int d^3x \,N\,\left(\frac{\pi^{ab}G_{abcd}\pi^{cd}}{\sqrt{|g|}}-(R-2\Lambda)\sqrt{|g|}\right)\\
   H(\xi)&=&\int d^3x\, \pi^{ab}(\mathcal L_\xi g)_{ab},
 \end{array}
\end{equation}
and CMC--gauge is obtained by imposing
\begin{equation}\label{equ:CMCgauge}
 \pi(x)=\langle \pi \rangle \sqrt{|g|}(x),
\end{equation}
as a gauge--fixing condition for the scalar ADM constraints $S(N)$, as introduced by York and collaborators \cite{York:1971hw,York:1972sj,York:1973ia,O'Murchadha:1974nc,O'Murchadha:1974nd}. The gauge condition implies that the trace $\pi/\sqrt{|g|}$ of the momenta is a spatial constant, which combined with the  vector constraints $H(\xi)$ implies that the momenta $\pi^{ab}$ are transverse and with constant trace. These conditions are preserved under spatial conformal transformations of the initial data $g_{ab},\pi^{ab}$, which can be used to solve the ADM scalar constraints by solving the Lichnerowicz-York equation
\begin{equation}\label{equ:LYE}
 8 \Delta \Omega =\left(\frac 3 8 \tau^2-2\Lambda\right)\Omega^5+R \,\Omega-\frac{\pi^{ab}_{TT}\pi^{TT}_{ab}}{|g|}\Omega^{-7},
\end{equation}
where $TT$ denotes the transverse traceless part and $\tau=\frac 2 3 \langle \pi\rangle/\sqrt{|g|}$ denotes York time. One can show that the positive solution $\Omega$ of (\ref{equ:LYE}) a compact Cauchy surface without boundary is unique if $\tau^2>\frac{16}{3}\Lambda$. The preservation of the gauge--fixing condition (\ref{equ:CMCgauge}) requires that the lapse $N_o$ satisfies the lapse--fixing equation $\{S(N_o),\pi(x)-\langle\pi\rangle\sqrt{|g|}(x)\}=0$, so the total CMC--Hamiltonian consists can be written as
\begin{equation}
 H_{CMC}=S(N_o)+H(\xi).
\end{equation}

\subsection{Construction of Shape Dynamics}

The CMC gauge--fixing condition (\ref{equ:CMCgauge}) generates spatial conformal transformations that preserve the total spatial volume; we can thus unfix ADM in CMC gauge to a gauge theory that has spatial diffeomorphism--invariance and local spatial conformal invariance \cite{Gomes:2010fh}. The simplest way \cite{Gomes:2011zi} to do this is by constructing a linking gauge theory through the canonical analogue of the procedure that I called ``best matching'' in section \ref{sec:Ontology}. For this we extend the ADM phase space with a scalar degree of freedom $\phi(x)$ and its canonically conjugate momentum density $\pi_\phi(x)$. The physical theory is unchanged, if we impose the additional first class constraints
\begin{equation}
 Q(\rho)=\int d^3x \,\rho\,\pi_\phi,
\end{equation}
which turn the phase space extension into pure gauge. We now implement best matching w.r.t. conformal transformations that preserve the total spatial volume through a canonical transformation generated by
\begin{equation}
 F=\int d^3x \left(\Pi^{ab}(e^{4\hat \phi}g_{ab})+\Pi_\phi \phi\right),
\end{equation}
where capital letters denote transformed variables and $\hat \phi:=\phi-\frac 1 6 \ln\langle e^{6\phi}\rangle$. The best--matching canonical transformation is
\begin{equation}\label{equ:BestmatchTrf}
 \begin{array}{rcl}
   g_{ab}&\to&e^{4\hat\phi}g_{ab}\\
   \pi^{ab}&\to&e^{-4\hat\phi}\left(\pi^{ab}-(1-e^{6\hat\phi})\langle\pi\rangle\sqrt{|g|}g^{ab}\right)\\
   \phi&\to&\phi\\
   \pi_\phi&\to&\pi_\phi-4\left(\pi-\langle\pi\rangle\sqrt{|g|}\right).
 \end{array}
\end{equation}
This transforms the constraints into
\begin{equation}
 \begin{array}{rcl}
  S(N)&\to&\int d^3x\,N\,\left(\frac{\sigma^a_b\sigma^b_a}{\sqrt{|g|}}e^{-6\hat \phi}+(2\Lambda-\frac 1 6 \langle \pi \rangle^2)\sqrt{|g|}e^{6\hat\phi}-\bar R \sqrt{|g|}e^{2\hat\phi}+...\right)\\
  H(\xi)&\to&\int d^3x\left(\pi^{ab}(\mathcal L_\xi g)_{ab}+\pi_\phi\mathcal L_\xi \phi+...\right)\\
  Q(\rho)&\to&\int d^3x\,\rho\,\left(\pi_\phi-4\left(\pi-\langle\pi\rangle\sqrt{|g|}\right)\right),
 \end{array}
\end{equation}
where $...$ denotes terms that vanish when the constraints $Q(\rho)$ hold and where $\bar R$ denotes the $e^{4\hat\phi}$ times the canonically transformed $R$ and where $\sigma^a_b$ denotes the trace--free part of the metric momenta. This linking theory can be partially gauge--fixed with $\phi\equiv 0$, which leads to the phase space reduction $(\phi,\pi_\phi)\to\left(0,4\left(\pi-\langle\pi\rangle\sqrt{|g|}\right)\right)$. This phase space reduction trivializes the constraints $Q(\rho)$ and recovers the ADM description of GR. 

The partial gauge--fixing $\pi_\phi\equiv 0$ constructs the Shape Dynamics description of GR. This construction is simplified if we replace $\hat \phi \to \phi$ and impose the volume preservation condition 
\begin{equation}\label{equ:VolPreserv}
\int d^3x \sqrt{|g|}\left(e^{6\phi}-1\right)=0.
\end{equation}
Imposing $\pi_\phi\equiv 0$ simplifies
\begin{equation}
   H(\xi) \,\to\, \int d^3x \,\pi^{ab}(\mathcal L_\xi g)_{ab},\,\,\,\,\,
   Q(\rho)\,\to\,(-1/4)\int d^3x\,\rho\,\left(\pi-\langle\pi\rangle\sqrt{|g|}\right),
\end{equation}
which means that spatial diffeomorphisms and spatial conformal transformations that preserve the total spatial volume are gauge transformations. The phase space reduction is determined by solving the scalar constraints for $\phi$, which using $\Omega:=e^\phi$ transforms into the Lichnerowicz-York equation (\ref{equ:LYE}), whose positive solution we denote by $\Omega_o[g,\pi;x)$. The phase space reduction is thus
\begin{equation}
 (\phi,\pi_\phi)\to (\ln(\Omega_o[g,\pi]),0).
\end{equation}
Inserting this into the volume preservation condition yields the constraint
\begin{equation}
  C=\int d^3x\,\sqrt{|g|}\left(1-e^{6\phi_o[g,\pi]}\right)\approx 0.
\end{equation}
We observe that this constraint takes the form of a time--reparametrization constraint. We deparametrize $C$ by fixing time to be equal to York time $\tau=\frac 3 2 \langle \pi\rangle$. This yields the physical Hamiltonian and conformal constraints
\begin{equation}
 H_{SD}=\int d^3x\sqrt{|g|}\,e^{6\phi_o[g,\pi;\tau]},\,\,\,\,Q(\rho)=\int d^3x\,\rho\,\left(\pi-\frac 2 3 \tau\right).
\end{equation}
We thus arrived at a description of gravitational dynamics that in which spatial diffeomorphisms and spatial conformal transformations are gauge--transformations \cite{Koslowski:2012uk}. We call this description of gravity Shape Dynamics, because it describes gravity as a dynamical theory of the spatial conformal geometry, or, for short, as a dynamical theory of the ``shape'' of space.

\subsection{A useful gauge condition for Shape Dynamics}\label{sec:usefulGauge}

The the construction of the Shape Dynamics Hamiltonian requires the solution of the Lichnerowicz-York equation, which practically very cumbersome. This situation can be simplified by the observation that the Lichnerowicz-York equation reduces to a polynomial if the gauge condition for the conformal factor is is chosen in such a way that the Lichnerowicz-York equation is solved by a spatial constant. Using $\tilde M(x):=M(x)-\langle M\rangle$ and $\tilde R(x):=R(x)-\langle R\rangle$ we can write the Lichnerowicz-York equation as
\begin{equation}
 8\Delta \Omega(x)=T\,\Omega^5(x)+(\tilde R(x)+\langle R\rangle)\Omega(x)+(\tilde M(x)+\langle M\rangle)\Omega^{-7}(x)
\end{equation}
Let us denote the positive solution of $\beta^3+\langle R\rangle/T \beta^2+\langle M\rangle/T=0$ by
\begin{equation}
 \alpha=\Delta-r+\frac{r^2}{\Delta},
\end{equation}
where $r=\langle R\rangle/T$, $m=\langle M\rangle/T$ and $\Delta=\sqrt[3]{\frac{1}{2}\left(\sqrt{27m \left(27 m+4 r^3\right)}-27m\right)-r^3}$. It follows that the spatial constant $\sqrt[4]{\alpha}$ solves the Lichnerowicz-York equation if the condition
\begin{equation}\label{equ:ADMgauge}
 M(x)=\alpha^2\,R(x)
\end{equation}
is satisfied, because the LHS and RHS then simplify to $8\Delta \sqrt[4]{\alpha}=0$ and $T\alpha^3+\langle R \rangle\alpha^2+\langle M\rangle=0$. Using the uniqueness of the positive solution to the Lichnerowicz-York equation, we conclude that the Shape Dynamics Hamiltonian turns into
\begin{equation}\label{equSimpleSDhamiltonian}
 H_{SD}=\int d^3x \sqrt{|g|}\alpha^{3/2}=\alpha^{3/2} V =V\left(\Delta-r+\frac{r^2}{\Delta}\right)^{3/2}
\end{equation}
if the condition (\ref{equ:ADMgauge}) is satisfied. Thus, if we choose CMC initial data and choose (\ref{equ:ADMgauge}) as a gauge condition, then we can evolve the system with the Hamiltonian (\ref{equSimpleSDhamiltonian}). Hence, if one finds a framework in which (\ref{equSimpleSDhamiltonian}) can be quantized before phase space reduction, say as $\hat H_o$, and if one can implement the solution to the conformal constraints through a projection operator $P$, which projects out the local conformal factor, then one would have a natural candidate for a quantum Hamiltonian in $P\,H_o\,P$. We will come back to this possibility in section \ref{sec:IshamQuant}.

\subsection{Coupling Shape Dynamics to Matter}

The Shape Dynamics description of a matter--gravity system can be constructed following precisely the same construction as in the pure gravity case and it works whenever the Lichnerowicz-York equation has a positive solution (this has been investigated in \cite{Isenberg:1976fq,Isenberg:1977ja,Isenberg:1977fs}). I will here only outline the procedure. The construction of the Shape Dynamics description of e.g. the Einstein-Maxwell system is straightforward \cite{Gomes:2011au}.

We denote the matter configuration degrees of freedom collectively by $\phi_A$ and their conjugate momentum densities by $\pi^B$. The ADM description of a generic gravity--matter system will have three sets of constraints
\begin{equation}
 \begin{array}{rcl}
   S(N)&=&S_{grav.}(N)+\int d^3x N\,\mathcal H_{matt.}(g_{ab},\phi_A,\pi^B)\\
   H(\xi)&=&H_{grav.}(\xi)+\int d^3x\,\pi^A(\mathcal L \phi)_A\\
   G(\lambda)&=&\int d^3x \sqrt{|g|}\, \lambda_\alpha\, G^\alpha(g_{ab},\phi_A,\pi^B),
 \end{array}
\end{equation}
where the subscript $grav.$ refers to the expressions for the pure gravity constraints, $\mathcal H_{matt.}$ denotes the matter Hamiltonian density and where the constraints $G(\lambda)$ collectively denote the matter constraints (such as the Gauss--constraints of Yang-Mills or Maxwell). To best match w.r.t. spatial Weyl--transformations that do not change the total volume, we introduce, as in the pure gravity case the auxiliary field $\phi$ and its canonically conjugate momentum density $\pi_\phi$ and impose the constraints $Q(x)=\pi_\phi(x)\approx 0$. The generating functional for the best--matching transformation is
\begin{equation}
 F=\int d^3x\left(\Pi^{ab}e^{4\hat \phi}g_{ab}+\phi\,\Pi_\phi+\phi_A\,\Pi^A\right),
\end{equation}
where Greek capitals denote the transformed fields. Notice that this transformation scales only gravity and leaves matter degrees of freedom unscaled, which means that the conformal transformation acts on geometry only. It follows that the canonical transformation acts in the gravity sector as equation (\ref{equ:BestmatchTrf}) and trivially on matter, i.e.
\begin{equation}
 \phi_A \to \phi_A,\,\,\,\,\,\pi^A\to\pi^A.
\end{equation}
It turns out that the Gauss--constraint densities are independent of gravitational variables, so the Gauss--constraints are transformed into themselves, while the diffeomorphism-- and $\pi_\phi$--constraints $Q(\rho)$ transform weakly into
\begin{equation} \label{equ:SDmattConstr}
 \begin{array}{rcl}
  G(\lambda)&\to&G(\lambda)\\
  Q(\rho)&\to&\int d^3x\,\rho\left(\pi_\phi-4\left(\pi-\langle\pi\rangle\sqrt{|g|}\right)\right)\\
  H(\xi)&\to&\int d^3x\,\left(\pi^{ab}(\mathcal L_\xi g)_{ab}+\pi_\phi \mathcal L_\xi \phi+\pi^A(\mathcal L_\xi \phi)_A\right).
 \end{array}
\end{equation}
We will, as in the pure gravity case, replace $\hat \phi \to \phi$ and impose the volume--preservation condition $\int d^3x \sqrt{|g|}(1-e^{6\phi})=0$ and denote $\Omega=e^\phi$. Using this notation, the ADM scalar constraints transform into the Lichnerowicz--York equation
\begin{equation}\label{equ:LYmatter}
  8 \Delta \Omega =\left(\frac 3 8 \tau^2-2\Lambda\right)\Omega^5+R \,\Omega-\frac{\pi^{ab}_{TT}\pi^{TT}_{ab}}{|g|}\Omega^{-7}-\frac{\mathcal H_{stand.}}{\sqrt{|g|}}\Omega^{-3}+...,
\end{equation}
where $\mathcal H_{stand.}$ stands for standard matter, such as minimally coupled massless Dirac--, pure Maxwell-- and Yang-Mills--Hamilton densities and where $...$ stands for additional terms, such as scalar field couplings and mass terms. For non--negative $\mathcal H_{stand.}$ and vanishing extra terms $...$ one finds that the positive solution $\Omega_o[g,\pi;\phi_A,\pi^A;x)$ to (\ref{equ:LYmatter}) is unique whenever $\tau>\frac{16}{3}\Lambda$. We can thus proceed as in pure gravity case and find the volume constraint $\int d^3x \sqrt{|g|}\left(\Omega_o^6[g,\pi;\phi_A,\pi^A]-1\right)$. Deparametrization in York--time then gives the physical Hamiltonian
\begin{equation}
 H_{SD}=\int d^3x\sqrt{|g|} \Omega_o^6[g,\pi;\phi_A,\pi^A]
\end{equation}
and the Shape Dynamics constraints (\ref{equ:SDmattConstr}). We see, that a gravity--matter system can be described through Shape Dynamics, which implies in particular that standard matter couples only to the degrees of freedom of the conformal 3-geometry. 

\subsection{Excursion: Doubly General Relativity}

We saw that ADM in CMC--gauge can be gauge--unfixed as either the ADM description of gravity or as the Shape Dynamics description, so the the BRST--gauge--fixed Hamiltonian can be constructed such that it is annihilated by two differentials: One that encodes the space--time gauge--symmetry of the ADM description and one that encodes the spatial conformal symmetry of the Shape Dynamics description \cite{Gomes:2012hh}. To obtain this description, we start with the BRST--construction for ADM. It requires the introduction of Grassmann--valued scalar ghosts $\eta(x)$ for the ADM scalar constraints $S(N)$ and Grassmann-valued vector ghosts $\eta^a(x)$ for the diffeomorphism constraints $H(\xi)$. The canonically conjugate ghost momentum densities are denoted by $P$ and $P_a$. The BRST generator for the ADM--description is
\begin{equation}
 \Omega_{ADM}=\int d^3x\left( S(\eta)+H(\vec \eta)+\eta^b\eta^a_{,b}P_a+\frac 1 2 \eta^a\eta_{,a}P+\eta\eta_{,c}p_bg^{bc}\right).
\end{equation}
This BRST--generator encodes spatial diffeomorphism--invariance and the refoliation--invaraince of the ADM description. There is a huge freedom how to construct the BRST--charge corresponding to spatial conformal invariance of Shape Dynamics: It has to include a term $\bar \Omega_{SD}=\int d^3x\,P\,\frac{\pi}{\sqrt{|g|}}+...$ which encodes spatial conformal invariance through the BRST--transform
\begin{equation}
 \delta_{SD-BRST} g_{ab}=\{\bar\Omega_{SD},g_{ab}\}=\frac{P}{\sqrt{|g|}}\,g_{ab}+...
\end{equation}
but there is a huge freedom in choosing the additional terms indicated by $...$. An interesting choice is to construct a conformally invariant gauge fixing of the spatial diffeomorphisms of the form $F^k(x)=(g^{ab}\delta^k_c+\frac 1 3 g^{ak}\delta^b_c)e^c_\alpha(\nabla_a-\hat\nabla_a)e^\alpha_b$, where the hat denotes the covariant derivative w.r.t. a fixed background metric with frame $e^a_\alpha$. With this choice, we find the BRST--charge for the conformal description as
\begin{equation}
 \bar \Omega_{SD}=\int d^3x\,\left(P\frac{\pi}{\sqrt{|g|}}+P_aF^a+\frac 1 2 \frac{P}{\sqrt{|g|}}P_a\eta^a\right),
\end{equation}
which is a charge that generates local Weyl--transformations and ultralocal transformations of the metric momenta, whose algebra over each point mimics the algebra of special conformal transformations. The gauge--fixed Hamiltonian 
\begin{equation}
 H_{BRST}=\{\Omega_{ADM},\bar \Omega_{SD}\}
\end{equation}
takes a very complicated form and I will only discuss the term that is important for the interpretation of this Hamiltonian as an evolution generator. This  is the ghost--free term 
\begin{equation}
 H_{BRST}=S(\frac{\pi}{\sqrt{|g|}})+H(\vec F)+\mathcal O(\eta).
\end{equation}
We see that this gauge--fixed Hamiltonian has a simple interpretation as an ADM generator of evolution with lapse $\frac{\pi}{\sqrt{|g|}}$ and shift $F^a$. We conclude this excursion with emphasizing that $H_{BRST}$ can be viewed as in two complementary ways: Either as a gauge fixing of the ADM description of gravity or as a gauge--fixing of a theory with local conformal gauge--symmetry.

\section{Example: Shape Dynamics on the 2+1 Torus}\label{sec:2+1}

Let us start the investigation of the implications of the Shape Dynamics picture by considering a simple example: Pure gravity on a torus in 2+1 dimensions allows us to perform many explicit derivations that are not feasible in more complicated models. I will focus on the ADM description of this system in constant mean (extrinsic) curvature (CMC) gauge, as it was first done by Moncrief in \cite{Moncrief:1989dx}. Using the spatial metric $g_{ab}$ and its canonically conjugate momentum density $\pi^{ab}$, we can write down the ADM system in terms of the initial value constraints
\begin{equation}
 \begin{array}{rcl}
  S(N)&=&\int d^2x\,N\left(\frac 1{\sqrt{|g|}}\pi^{ab}(g_{ac}g_{bd}-g_{ab}g_{cd})\pi^{cd}-(R-2\Lambda)\sqrt{|g|}\right)\\
  H(v)&=&\int d^2x\,\pi^{ab}(\mathcal L_v g)_{ab},
 \end{array}
\end{equation}
where $S(N)$ denotes the Hamilton constraint smeared with lapse $N$ and $H(v)$ the diffeomorphism constraint smeared with shift $v$. The constraints $S(N)$ play of course a dual role, as they are not only initial value constraints, but also generators of dynamics, i.e. the constraints generate evolution in time.

\subsection{Description on Reduced Phase space}

Let us outline the construction of the York-Hamiltonian for the 2+1 torus, which is based on the 2+1--dimensional CMC gauge fixing condition for the scalar constraints, which states that York time
\begin{equation}
 \tau=g_{ab}\pi^{ab}/\sqrt{|g|}
\end{equation}
is a spatial constant. Finding a complete gauge-fixing of the spatial diffeomorphisms is more complicated. Fortunately, Teichm\"uller theory tells us that the space of conformal geometries on the 2-torus is 2-dimensional. Using a global chart $[0,1)^2$ for the torus\footnote{To simplify the notation, I present the treatment of the diffeomorphisms in a preferred global chart. One could of course write down the same equations in a covariant way using coordinate functions, frame fields and Lie derivatives.}, we can write the components of the conformal metric $\rho_{ab}=g_{ab}/\sqrt{|g|}$ in this chart as
$$
  \rho_{ab}=\left(\begin{array}{cc}\alpha&\beta\\b&\frac{1+\beta^2}{\alpha}\end{array}\right).
$$
Imposing $\sum_b \rho_{1b,b}(x_1,x_2)=0$ and $\rho_{21,1}(x_1,x_2)=0$ as well as $\rho_{11,2}(0,x_2)=0$, turns out to completely gauge-fix the diffeomorphisms. This fixes $\alpha,\beta$ to be spatial constants. We now follow the York procedure and solve the diffeomorphism constraint by setting the longitudinal part of $\pi^{ab}$ to vanish. Using $g_{ab}=e^{2\lambda} \rho_{ab} \omega=h_{ab}e^{2\lambda}$, where $\omega=dx^1\wedge dx^2$, allows us to write the scalar constraints as the Lichnerowicz-York equation
\begin{equation}
  \Delta_h \lambda = (\frac 1 4 \tau^2-\Lambda)e^{2\lambda}-\frac{\pi^{ab}_{TT}h_{ac}h_{bd}\pi^{cd}_{TT}}{2\sqrt{|h|}}e^{-2\lambda}.
\end{equation}
We can parametrize the TT-part of the metric momenta in the global chart explicitly with homogeneous components (see e.g. \cite{Carlip:2004ba}):
$$
  \pi^{ab}_{TT}=\frac 1 2 \left(\begin{array}{cc} (\tau_1^2+\tau_2^2)p_2-2\tau_1\tau_2p_1&\tau_2p_1-\tau_1p_2\\ \tau_2p_1-\tau_1p_2&p_2\end{array}\right),
$$
where the parameters $p_1,p_2$ constructed to be canonically conjugate to the Teichm\"uller parameters $\tau_1,\tau_2$, which parametrize the gauge-fixed metric as $h_{ab}=\tau_2^{-1}\left(dx_1\otimes dx_1+\tau_1(dx_1\otimes dx_2+dx_2\otimes dx_1)+(\tau_1^2+\tau_2^2)dx_2\otimes dx_2\right)$. All coefficients of the Lichnerowicz-York equation are homogeneous in this parametrization. This implies that the Lichnerowicz-York equation is solved by the homogeneous solution
$ e^{4\lambda_o}=\tau_2^2\frac{p_1^2+p_2^2}{\tau^2-4\Lambda}$. The generator of evolution in York-time is the on-shell volume $\int d^2x \sqrt{|h|} e^{2\lambda_o}$. It reads in our parametrization
\begin{equation}
 H_{\small\textrm{York}}=\tau_2\sqrt{\frac{p_1^2+p_2^2}{\tau^2-4\Lambda}}.
\end{equation}
So far we have just followed the standard construction of the York Hamiltonian on reduced phase space. The simplicity of the model allows us to construct Shape Dynamics directly by reversing the role of the Hamilton constraints and the CMC-gauge condition.

\subsection{Reinterpretation in terms of Shape Dynamics}

The simplicity of the 2+1 torus allows us to discover Shape Dynamics in the CMC reduced phase space formulation, rather than to construct it as we did in 3+1 dimensions. To do this, we need two observations about the CMC formulation:
\begin{enumerate}
 \item The CMC-reduced phase space can be identified with the gauge orbits of conformal diffeomorphisms. After all, Teichm\"uller space, the reduced configuration space, can be obtained as the equivalence classes of spatial metrics modulus diffeomorphisms\footnote{We will be mostly concerned with small diffeomorphisms, i.e. those diffeomorphisms that are connected to the identity diffeomorphism. Invariance under large diffeomorphisms, or modular invariance, is however preserved.} and conformal transformations.
 \item There is a simple diffeomorphism- and Weyl-invariant phase space function, whose restriction to reduced phase space coincides with the York Hamiltonian:
 \begin{equation}
   V_o=\int d^2x \sqrt{\frac{\pi^{ab}(g_{ac}g_{bd}-\frac 1 2 g_{ab}g_{cd})\pi^{cd}}{\frac 1 2 \tau^2-2\Lambda}}.
 \end{equation}
\end{enumerate}
This suggests to consider the following dynamical system with first class constraints
\begin{equation}\label{equ:2+1constraints}
 \begin{array}{rcl}
   H(v)&=&\int d^2x \,\pi^{ab}(\mathcal L_v g)_{ab}\\
   D(\rho)&=&\int d^2x\,\rho\,g_{ab}\pi^{ab},
 \end{array}
\end{equation}
and the time-dependent Hamiltonian
\begin{equation}
 \tilde H_{\small\textrm{York}}=V_o=\int d^2x \sqrt{\frac{\pi^{ab}(g_{ac}g_{bd}-\frac 1 2 g_{ab}g_{cd})\pi^{cd}}{\frac 1 2 \tau^2-2\Lambda}}.
\end{equation}
The constraints $H(v)$ are the familiar spatial diffeomorphism constraints, while the constraints $D(\rho)$ generate spatial conformal (or Weyl-) transformations:
\begin{equation}
 \delta g_{ab}=\{g_{ab},D(\rho)\}=\rho \,g_{ab},\,\,\,\,\,\delta \pi^{ab}=\{\pi^{ab},D(\rho)\}=-\rho \,\pi^{ab}.
\end{equation}
The first property can be explicitly verified by imposing $R[g;x)=0, V=\int d^2x \sqrt{|g|}=1$ as a gauge condition for the conformal transformations and gauge-fixing the diffeomorphisms as we did in the previous subsection. Direct construction then shows that the reduced phase space is conveniently parametrized by the aforementioned Teichm\"uller parameters $\tau_1,\tau_2$ and their canonically conjugate momenta $p_1,p_2$.

To verify the second property, we first notice that $\tilde H_{\small\textrm{York}}$ is diffeomorphism invariant, because it is a density of weight one integrated over a compact space without boundary and depends solely on $g_{ab},\pi^{ab}$. Conformal invariance can be verified by checking the Poisson bracket $\{\tilde H_{\small\textrm{York}},D(\rho)\}=0$. Direct insertion of the parametrization of reduced phase space then verifies that $\tilde H_{\small\textrm{York}}$ coincides with $H_{\small\textrm{York}}$ on reduced phase space. We thus have a gauge theory of spatial diffeomorphisms and spatial Weyl-transformations that can be gauge-fixed in such a way that it completely coincides with the ADM-description of gravity in CMC gauge. The observables of this gauge theory can be identified with smooth functions of $\tau_1,\tau_2,p_1,p_2$. The second property then implies that the evolution of observables coincides with the CMC evolution of ADM observables\footnote{Dirac observables of the ADM system are 
usually required to Poisson commute with all Hamilton constraints. CMC evolution of ADM observables is then defined as the York-time dependence of Dirac observables when CMC gauge is imposed.}. This means that all observable predictions of the conformal gauge theory coincide with predictions for ADM observables. We can thus say conformal gauge theory describes the same classical physics as ADM gravity, since all objective predictions about the system must be stated in terms of observables. Moreover, given any solution $(\tau_1,\tau_2,p_1,p_2)(\tau)$ of the conofrmal gauge theory, we find the equivalent CMC solution by reinterpreting $\tau=\frac{g_{ab}\pi^{ab}}{\sqrt{|g|}}$ and setting the volume to its on-shell value $V=V_o(\tau_1,\tau_2,p_1,p_2;\tau)$.

The 2+1 torus allows for another simplification: So far, we have simply used the York-Hamiltonian, which depends on a dimensionful external time variable $\tau$. This time-dependence is however only an overall multiplicative factor $(\frac 1 2 \tau^2-2\Lambda)^{-\frac 1 2}$ and can thus be removed by a reparametrization of the time variable with global lapse $\mathcal N=\sqrt{\frac 1 2 \tau^2-2\Lambda}$. The relation of the new time variable $t$ and York time $\tau$ is given by
\begin{equation}
 t=\ln\left(\frac{\tau+\textrm{sgn}(\tau)\sqrt{\tau^2-16 \Lambda \tau_o^2}}{2 \tau_o}\right),
\end{equation}
where $\tau_o$ denotes an arbitrary, but fixed initial value of York time. This reparametrization yields the time-independent shape dynamics Hamiltonian
\begin{equation}\label{equ:2+1Hamiltonian}
 H_{SD}=\int d^2x\sqrt{\pi^{ab}(g_{ac}g_{bd}-\frac 1 2 g_{ab}g_{cd})\pi^{cd}}.
\end{equation}
This Hamiltonian generates an autonomous dynamics of shapes, i.e. a time-independent dynamics on Teichm\"uller space. This dynamics is a geodesic dynamics on Teichm\"uller space. It is thus convenient to define Shape Dynamics on the 2+1 torus as a gauge theory with the first class constraints (\ref{equ:2+1constraints}) and Hamiltonian (\ref{equ:2+1Hamiltonian}). 

\subsection{Effective Field Theory}

Gauge symmetries play an important role in effective field theory, which we will outline in the next section. The simplicity of the 2+1 torus allows a more direct approach based on a particular interpretation of causal dynamical triangulations (CDT).

Causal dynamical triangulations are often presented as a discretized path integral approach to the Wick-rotated Einstein-Hilbert action. It posses a preferred foliation, which provides the time variable that is thought to be Wick rotated. CDT is however, at the most basic level, a statistical mechanical model of spacetime geometry with a preferred foliation. The hypothesis underlaying CDT is that the continuum limit of the model contains a phase that is effectively described by General Relativity or Ho\v{r}ava--Lifshitz gravity. Analytic results pertaining to CDT in more than 1+1 dimensions are limited, but CDT is successfully tested using numerical simulations.

Recently, CDT simulations of the 2+1 torus have tested the evolution of volume and shape degrees of freedom \cite{Budd:2011zm,Budd:thesis}. A quick summary of the results is as follows:
\begin{enumerate}
 \item The average trajectory in Teichm\"uller space is numerically consistent with a continuum evolution generated by $H_{SD}$.
 \item The average evolution of the volume is not consistent general covariance. This is related to the wrong sign in the kinetic term of the conformal mode.
\end{enumerate}
At first sight, the second result seems to suggest that the CDT simulations favor a non-covariant generalization of General Relativity, such as Ho\v{r}ava--Lifshitz gravity. However, if we interpret the preferred foliation as York time slicing and adapt the shape dynamics interpretation that the conformal mode is pure gauge, then the results are completely consistent with gravity. Moreover, the preferred foliation in CDT seems to coincide York time slicing and thus to coincide with the preferred foliation of shape dynamics. One can thus argue that CDT dynamically selects the Shape Dynamics description of gravity.

Notice, that the conformal mode problem is completely avoided in Shape Dynamics. The conformal factor is pure gauge in Shape Dynamics. The only place where the conformal mode plays a role is in the reinterpretation of a Shape Dynamics trajectory in terms of spacetime geometry, because the translation into the spacetime picture requires that we pick the conformal factor that solves the Lichnerowicz--York equation. 

It is definitely premature to call this observation evidence for the conjecture that CDT data should be interpreted as Shape Dynamics trajectories, because pure gravity on the torus in 2+1 dimensions is a very special dynamical system. It is nevertheless an interesting example that illustrates the fact that the spatial conformal ontology of Shape Dynamics and the spacetime ontology of General Relativity may yield different effective field theories by interpreting the predictions of a model differently.

\section{Effective Filed Theory}\label{sec:effective}

We saw that Shape Dynamics provides a description of gravitational dynamics as dynamical theory of the spatial conformal geometry, but, so far, the emphasis has been on the fact that classical Shape Dynamics is, at least locally, indistinguishable from the ADM description. This mere change in view--point, although it may aid the investigation of certain questions about classical gravity, may be of very limited interest for the discussion of classical gravity, because one can in principle work out any problem in classical gravity using the ADM description. 

I will argue now that this shift in view--point may provide valuable insights which could help the systematic search for quantum gravity. The two obvious points are that the Shape Dynamics description solves two very difficult and entangled problems of canonical quantum gravity and replaces them with one systematically attackable problem: These problems are (1) the problem of gauge--invariance is simplified by identifying the conformal geometry as the physical components of the gravitational field and (2) the problem of time is solved, since the Shape Dynamics description posses a physical Hamiltonian. The condensed problem of quantizing the Shape Dynamics description is to find a quantum operator that generates a dynamics that can be identified with the dynamics generated by the Shape Dynamics Hamiltonian in a sensible semiclassical limit. 

Before I will come back to how this problem may be attacked in the Loop Quantum Gravity and polymer quantization framework, I will consider the standard effective field theory interpretation of te fact that gravity can be described through Shape Dynamics.

\subsection{Shape Dynamics theory space}\label{sec:TheorySpace}

Path--integrals can be used to define quantum field theories, but this definition is, with very few exceptions, only formal, because the path--integral itself is, with very few exceptions, ill-defined and requires regularization (in the form of a cut-off at a high--energy scale $\Lambda$). The subsequent removal of the regulator (i.e. the limit $\Lambda \to \infty$) requires in general renormalization ``flow'' of the action. The modern view of quantum field theory is based on the Kadanoff-Wilson picture of renormalization and describes quantum field theories a renormalization--group trajectory of flowing effective actions (for a review see appendix A in \cite{Niedermaier:2006wt}). A fundamental quantum field theory is, in the Kadanoff-Wilson picture, a complete renormalization group trajectory, i.e. a one--parameter family of effective actions that can be extended to all values of the cut--off $\Lambda$, so the cut--off can be removed. 

A very important concept in this framework is the so--called ``theory space,'' which is the space of action functionals on which the renormalization group equation is defined. A simple explanation of ``theory space'' can be given in the exact renormalization group framework, which I will briefly explain using the renormalization group equation that is satisfied by the effective average action $\Gamma_k[\phi]$, which is an interpolation between a bare action $S=\Gamma_\Lambda$, in the limit $\Lambda\to \infty$, and the generating functional for the one--particle irreducible Green's functions $\Gamma[\phi]$ in the limit $k\to 0$. Its renormalization group flow is governed by Wetterich's equation
\begin{equation}\label{equ:Wetterich}
 k\partial k \Gamma_k[\phi]=\frac 1 2 STr\left(\frac{k\partial k R_k}{\Gamma_k^{(2)}+R_k}\right),
\end{equation}
where $\phi$ runs over the fields over which the path integral performed and $R_k$ denotes a scale--dependent mass terms, which gives modes with wavelength longer than $k^{-1}$ an effective mass of order $k^2$ and vanishes effectively for short wavelength modes and where $\Gamma_k^{(2)}$ denotes the second variation of $\Gamma_k$ w.r.t. $\phi$. We can thus replace the evaluation of path--integrals with finding solutions $\Gamma_k$ to (\ref{equ:Wetterich}) with initial condition $S=\Gamma_\Lambda$ at $k=\Lambda$ and evaluating $\Gamma_{k=0}$. In other words: the definition of the path integral can be entirely based on the renormalization group equation (\ref{equ:Wetterich}) of effective field theories, one never needs a path--integral. Moreover, the renormalization group allows one to study effective field theories without ever using an initial condition $S$; instead the logic is usually reversed and the renormalization group equation is used as a tool in the search for critical bare actions $S$.

Using this renormalization--group approach reveals the first ingredient for the theory space: the left and right--hand side of (\ref{equ:Wetterich}) are functionals of the fields $\phi$. Moreover, all physical symmetry requirements, which when imposed on $\Gamma_k$ are implied on the RHS of (\ref{equ:Wetterich}) are preserved by the renormalization group flow. Hence, the theory space contains functionals of the fields $\phi$ that are invariant under these symmetries.

It is in general not possible to choose the regulator term $R_k$ to preserve gauge--symmetry, so gauge--invariance can not be directly imposed on theory space. Instead, gauge--invariance implies complicated (modified) BRS Ward--identities $\mathcal W_k (\Gamma_k)=0$, which acquire modification terms\footnote{These modification terms are important for the discussion of anomalies: They tell us how non--gauge symmetric $\Gamma_\Lambda$ has to be chosen to ensure that the effective action $\Gamma_0$ has gauge--symmetry. If one chooses the bare action $S=\lim_{\Lambda\to\infty}\Gamma_\Lambda$ to have gauge symmetry then one violates in general $\mathcal W_\Lambda(\Gamma_\Lambda)=0$ and the effective action $\Gamma$ will posses anomalous terms.} due to the transformation properties of the regulator\footnote{The Slavnov-Taylor--identities are also complicated by the transformation properties of the regulator.}. However, it follows from the structure of (\ref{equ:Wetterich}) that the modified Ward--identities flow 
as
\begin{equation}
 k\partial_k \mathcal W_k=-\frac 1 2 G_k^{AB}(k\partial_k R_k^{AB})G_k^{CD}\frac{\delta^2 \mathcal W_k}{\delta \phi^B\,\delta\phi^D},
\end{equation}
where $A,B,C,D$ denote DeWitt--indices and where $G_k=(\Gamma_k^{(2)}+R_k)^{-1}$ denotes the full (field--dependent) propagator (for a review see \cite{Gies:2006wv}). The flow equation for the modified Ward--identities tells us that the modified Ward--identities $\mathcal W_k(\Gamma_k)=0$ are preserved by RG-flow. This reveals the second ingredient for the theory space: The space of effective action functionals is restricted to the solutions to the modified Ward--identities. The usual effective action $\Gamma=\Gamma_{k=0}$ satisfies $\mathcal W_{k=0}(\Gamma)=0$. In a semiclassical limit, this equation turns into the statement that $\Gamma$ is BRST--invariant plus $\mathcal O(\hbar)$ corrections.

We will now assume that we use this effective field theory framework for quantum Shape Dynamics. For this we do not assume a bare action and thus not a bare Hamiltonian. Instead, we assume the renormalization group equation (\ref{equ:Wetterich}) and the gauge symmetries of Shape Dynamics. We encode these in the BRST--charge
\begin{equation}
 \Omega_{SD}=\int d^3x\left(\eta \pi+\eta^a g_{ac}{\pi^{cd}}_{;d}+\eta^b{\eta^a}_{,b}P_a+\frac 1 2 \eta^a\eta_{,a}P\right).
\end{equation}
Assuming a BRST--invariant bare measure, one can assume an explicit form for the regulator and use this charge and the to derive a modified Ward--identity $\mathcal W^{SD}_k(\Gamma_k)=0$. We now solve the path--integral by evolving an initial condition $\Gamma_\Lambda=S$ with the renormalization group equation to the standard effective action $\Gamma$ at $k=0$. Since the regulator term vanishes at $k=0$, it follows that $\Gamma$ satisfies a standard Ward--identity $\mathcal W^{SD}(\Gamma)=0$, which is stating that the vacuum expectation value of a BRS variation vanishes. In a semiclassical regime, where the vacuum expectation values can be replaced with the field values plus corrections of order $\hbar$, we thus find
\begin{equation}
 \delta^{SD}_{BRS}\,\Gamma = \mathcal O(\hbar).
\end{equation}
I.e. in a semiclassical limit, the effective action is invariant under the BRST--transformations $\delta^{SD}_{BRST}\,f=\{f,\Omega_{SD}\}$. This means that the semicalssical  limit of a quantum Shape Dynamics theory posses the BRST--manifestation of spatial diffeomorphism-- and spatial Weyl--symmetry if it is constructed in the way I described here. Notice, that I did not use the standard path--integral prescription based in a BRST--invariant bare action, but that I required an initial condition $\Gamma_\Lambda$ for the renormalization group evolution that satisfies a {\it modified} Ward--identity, which ensures that the standard Ward--identity $\mathcal W^{SD}(\Gamma)$ holds for the effective action $\Gamma$.

The renormalization group equation (\ref{equ:Wetterich}) can be used as a tool in the systematic search for asymptotically safe theories, which are candidates for fundamental field theories. The notion of asymptotic safety was introduced by Weinberg in \cite{Weinberg:1979as} as a generalization of asymptotic freedom in the context of the search for quantum gravity. He called a (1) fundamental and (2) predictive field theory asymptotically safe. In the renormalization group approach these conditions are satisfied by a fixed point of the renormalization group with a finite dimensional critical manifold. This is because a renormalization group trajectory can be extended to arbitrary cut--off values $\Lambda$ if it ends in a fixed point, so it is fundamental. Moreover, the number of coupling constants that are relevant IR--deformations of the fixed point action is the dimension of the critical manifold, so a finite dimensional critical manifold implies a predictive theory.

The search for a physically acceptable gravitational fixed point action is an ongoing research program and is currently based on a theory space whose field content contains the spacetime metric and whose gauge--symmetries are spacetime diffemorphisms. The precise definition of this theory space is very important in the search for a fixed point action. 

Shape Dynamics posses a non--local Hamiltonian and a different theory space; its field content is the spatial metric and its gauge--symmetries are spatial conformal diffeomorphisms. A systematic search for a physically acceptable fixed point in the Shape Dynamics theory space has not been performed so far and it could be that the fixed point structure changes significantly due to the different structure of the theory space. 

\subsection{Loop Quantization of SD}

The most advanced canonical approach to quantum gravity is Loop Quantum Gravity \cite{Rovelli:2004tv,Thiemann:2007zz}. I call Loop Quantum Gravity most advanced, because it starts with a mathematically rigorous Hilbert space and representation of the kinematic constraints. A major open problem in the Loop Quantization program is finding a physically acceptable quantization of the Hamilton constraints for pure gravity. This is in part due to the fact that the canonical action of the scalar constraints does, unlike the kinematic constraints, not admit a simple geometric interpretation of the generated gauge--transformations. These problematic scalar constraints are replaced by spatial conformal constraints in the Shape Dynamics formulation of gravity, which posses a simple geometric interpretation. 

Let us start the Loop Quantization program with considering classical Shape Dynamics in Ashtekar variables. We start with extrinsic curvature variables, consisting of densitized inverse triads $E^a_i$ and extrinsic curvature $K^i_a$. These are related to the metric variables trough
\begin{equation}\label{equ:extrinsicVarDefi}
  g_{ab}=|E|E^i_aE^j_b\delta_{ij},\,\,\,\,\,\pi^{ab}=\frac{E^a_kE^d_l\delta^{kl}}{2|E|}\left(K^i_c\delta^b_d-K^i_d\delta^b_c\right)E^c_i. 
\end{equation}
This description has in addition to the scalar and diffeomorphism constraints also internal rotation constraints $G(\lambda)$; the full set of constraints for the Shape Dynamics description of gravity in these variables is
\begin{equation}
 \begin{array}{rcl}
   G(\lambda)&=&\int d^3x\, \lambda^i\, {\epsilon_{ij}}^kK^j_aE^a_k\\
   H(v)&=&\int d^3x\,K^a_a(\mathcal L_v E)^a_i\\
   C(\rho)&=&\int d^3x\,\rho\left(K^i_aE^a_i-\frac 2 3 \tau \sqrt{|E|}\right),
 \end{array}
\end{equation}
where $\tau$ denotes York time. Let us denote the solution to the Lichnerowicz-York equation (\ref{equ:LYE}) by $\Omega_o[K,E;x)$, where we insert the definition (\ref{equ:extrinsicVarDefi}) into (\ref{equ:LYE}). This allows us to write the physical Hamiltonian as the conformal York Hamiltonian as
\begin{equation}
 H_{SD}=\int d^3x\, \sqrt{|E|}\,\Omega_o^6[K,E].
\end{equation}
We will now use a canonical transformation generated by the generating functional
\begin{equation}
 F=\int d^3x\left(\beta K^i_a\tilde E^a_i+\tilde E^a_i\Gamma^i_a(\tilde E)\right),
\end{equation}
where tilde denotes the transformed variable, $\beta$ the Immirzi parameter and $\Gamma^i_a(E)$ the spin connection expressed as a function of the densitized inverse triad, to transform this system to Ashtekar variables. This leads to the canonical transformation
\begin{equation}
 \tilde A_a^i=\beta K^i_a+\Gamma^i_a(\tilde E),\,\,\,\,\,\,\,\tilde E^j_b=\frac{E^j_b}{\beta},
\end{equation}
where $\tilde A_a^i$ denotes the components of the Ashtekar connection and $\tilde E^j_b$ the rescaled densitized inverse triad\footnote{Notice that many authors do not perform this canonical transformation, but use the unscaled densitized inverse triad, which leads to a rescaling of the canonical Poisson bracket by a factor of $\beta$.}. After integration by parts, one finds that the diffeomorphism constraint takes a simple form and that the internal rotation constraint $G(\lambda)$ turns into the Gauss constraint for the Ashtekar connection:
\begin{equation}
 \begin{array}{rcl}
   G(\lambda)&=&\int d^3x\,\lambda^i \tilde D_a \tilde E^a_i\\
   H(v)&=&\int d^3x\,\tilde A^i_a(\mathcal L_v \tilde E)^a_i,
 \end{array}
\end{equation}
where $\tilde D_a$ denotes the Ashtekar covariant derivative, so $\tilde D_a E^a_i=\tilde E^a_{i,a}+\epsilon_{ijk}\tilde A^j_a\tilde E^a_k$. Unfortunately, the conformal constraints take a very complicated form after the canonical transformation
\begin{equation}
 C(\rho)=\int d^3x\,\rho\,\left((\tilde A^i_a-\Gamma^i_a(\tilde E))\tilde E^a_i-\frac{2}{3}\tau\beta^{-3/2}\sqrt{|\tilde E|}\right).
\end{equation}
This expression depends nonlinearly on the $\tilde E^a_i$, which are the Ashtekar momentum variables, which points to a serious challenge in the quantization of the conformal constraints. A direct calculation reveals the conformal transformation properties of the Ashtekar connection
\begin{equation}\label{equ:AshtekarConformal}
 \tilde A^i_a \to e^{-4\phi}\tilde A^i_a+(1-e^{-4\phi}\tilde \Gamma^i_a(\tilde E)+2\epsilon^{ijk}(\tilde E^{-1})^j_a\tilde E^b_k\phi_{,b},
\end{equation}
which explicitly shows that the conformal transformations can not be quantized as vector fields on the Ashtekar configuration space (i.e. the quantum completion of the space of field configurations of the Ashtekar connection). 

I will now describe an attempt to Loop Quantize the Shape Dynamics description of gravity; a more detailed account which will soon appear. This program consists of three steps.

{\bf Step 1: Kinematic Loop Quantization} This first is precisely what is done in textbook Loop Quantum Gravity. I will not repeat the entire program here, but only summarize essential features of this program that are important for the presentation here:
\begin{enumerate}
 \item The kinematic Hilbert space of Loop Quantum gravity is the space of gauge- and diffeomorphism-invraint functionals of the Ashtekar connection, that are square--integrable w.r.t. a group average obtained from the Ashtekar-Isham-Lewandowski measure. This Hilbert space has a basis given by diffeomorphism classes of spin--network functions of the Ashtekar connection.
 \item The standard interpretation of geometric operators provides the following geometric interpretation of the spin--network functions: The spin labels on edges of the spin--network represent quanta of area for transversing surfaces and the intertwiner labels at the vertices of the spin--network functions represent quanta of volume for surrounding regions and quanta of angles between surfaces that intersect at the vertex. 
 \item Spin--network functions are in a sense eigenfunctions of spatial geometry. More precisely: specifying a complete set of allowed area and volume quantum numbers and additionally allowed angle quantum numbers for a maximal commuting set of angle operators completely determines a spin--network function that is a simultaneous eigenfunction of the area- volume- and angle- operators with the respective allowed eigenvalues.
 \item One important fact is that angle operators do not commute with each other; their algebra does not even close among themselves, because the commutator of two angle operators contains in general terms proportional to a volume operator. This is related to the choice of recoupling basis.
\end{enumerate}

{\bf Step 2: Conformal Moduli} The conformal transformation properties of the Ashtekar connection (\ref{equ:AshtekarConformal}) makes it hopeless to quantize the generators of conformal transformations directly. Fortunately the situation is saved by geometry: On the one hand there is a clear interpretation of spatial conformal transformations in terms of spatial geometry. Loop Quantum Gravity on the other hand provides a standard interpretation of complete basis (the spin--network functions) in terms of spatial geometry. 

We observe classically that conformal transformations change any non--vanishing area to any other non--vanishing area and any non--vanishing volume to any other non--vanishing volume, while vanishing areas and volumes and all local angles are left invariant by conformal transformations. Moreover, two classical geometries, that agree for all angles and differ only for some non--vanishing areas and volumes, are representatives of the same conformal geometry. These two classical geometries are thus identified with the same classical state in Shape Dynamics.

The quantum analogue of this classical statement is to identify quantum states that agree in all quantum numbers except for some non--vanishing area and volume quantum numbers as the same physical state. This suggests to solve the quantum conformal constraints by quotienting this equivalence relation. We sketch the program as follows: 
\begin{enumerate}
 \item We use that spin--networks on different graphs are orthogonal, so we can consider each diffeomorphism class graphs separately. On each graph, we denote the number of edges by $n$ and the number of vertices by $m$, we choose a recoupling basis that diagonalizes the $m$ distinct volume operators. The spin--network functions are now given by
 \begin{equation}
   \left|[\gamma];j_1,...,j_n;V_1,...V_m;k_1,...k_l\right\rangle,
 \end{equation}
 where $[\gamma]$ denotes the diffeomorphism orbit of a graph $\gamma$, $j_1,...,j_n$ denotes the $n$ distinct area quantum numbers, $V_1,...,V_m$ denotes the $m$ distinct volume quantum numbers and where $k_1,..,k_l$ denotes the intertwiner labels that are independent of the volume\footnote{An additional redundancy, which I do not consider here, in this representation is due to graph automorphisms. The solution to the diffeomorphism constraint requires additional permutation symmetries between the labels $j_1,...,j_n;V_1,...V_m;k_1,...k_l$ when $[\gamma]$ has a nontrivial automorphism group.}. 
 \item Two states that disagree on any quantum label $[\gamma];j_1,...,j_n;V_1,...V_m;k_1,...k_l$ are orthogonal in the Loop Quantum Gravity Hilbert space. We can therefore build equivalence classes of the form
 \begin{equation}
   \left|[\gamma];j_1,...,j_n;V_1,...V_m;k_1,...k_l\right\rangle \sim \left|[\gamma];j^\prime_1,...,j\prime_n;V\prime_1,...V\prime_m;k_1,...k_l\right\rangle,
 \end{equation}
 where we require $j_i^\prime=0$ iff $j_i=0$ and $V_r^\prime=0$ if $V_r=0$. Orthogonality allows us to induce an inner product for the Loop Quantum Gravity inner product by fixing a unique representative $\left|[\gamma];j_1,...,j_n;V_1,...V_m;k_1,...k_l\right\rangle$ in each equivalence class and using this unique representative in place of the elements of the equivalence class. 
 \item For the procedure to be well defined, we need to construct a commuting set of operators $K_1,...,K_l$ that commute with all volume operators for each $[\gamma]$. I can construct these operators for simple $[\gamma]$, but I have not found a general construction prescription for this set of operators yet.
\end{enumerate}

{\bf Step 3: Effective Dynamics} So far, I presented a program for the construction of the physical Hilbert space for pure Loop Quantum Shape Dynamics. This by itself constitutes progress compared to the standard description of Loop Quantum Gravity. The implementation of a physically acceptable dynamics is however a wide open problem. This is due to the complicated non--local form of the Shape Dynamics Hamiltonian, which obstructs direct quantization efforts. This suggests to take an effective field theory approach and to perform a systematic search. For this one needs investigate the semiclassical behavior of trial Hamiltonians and refine the trial Hamiltonians until a physically acceptable candidate is found. 

\subsection{Opportunities for non--standard quantization programs}\label{sec:IshamQuant}

The reason for the successful kinematic quantization of Loop Quantum Gravity can be traced to the use of a non--Fock representation of a non--canonical commutation relations constructed from a diffeomorphism--invariant vacuum vector. In light of this success, one expects that much progress in the quantization of Shape Dynamics could be made if one where able to construct a non--Fock representation based on vacuum vector that is diffeomorphism and conformally invariant. 

To illustrate this possibility, I will describe the essential features of the operator algebra and GNS--functional used in Loop Quantization. 
\begin{enumerate}
 \item The operator algebra $\mathfrak A_o$ that underlies Loop Quantum Gravity is a completion of the algebra of finite sums of ordered pairs of the form
  \begin{equation}
    a =\sum_i f_i(A) G_i,
  \end{equation}
  where the functions $f_i(A)$ are continuous functions on the quantum configuration space (i.e. a particular topological completion of the space of Ashtekar connections) and the $G_i$ are elements of the group of Weyl--transformations on the quantum configuration space. The product is defined as the pointwise product  between the $f_i(A)$ and the group product between the $G_i$ and the non--canonical commutation relations $G_i^{-1} f(A) G_i:=f(T_i A)$, where $T_i$ denotes the action of a Weyl--transformation on the quantum configuration space. The involution is defined as complex conjugation of the $f_i(A)$ and group inversion of the $G_i$.
 \item A linear functional on the operator algebra generated by the finite sums $a\in\mathfrak A_o$ is defined through
  \begin{equation}
    \omega(a)=\sum_i \int d\mu_{AIL}(A) f_i(A),
  \end{equation}
  where $d\mu_{AIL}(A)$ denotes the Ashtekar-Isham-Lewandowski measure. It can be shown that $\omega(a^*a)\ge 0$ for all $a\in\mathfrak A_o$, so $\omega$ can be used to perform the GNS construction.
  \item The GNS--representation $\pi:\mathfrak A_o\to B(\mathcal H)$ on the GNS--Hilbert space $\mathcal H$ constructed form $\omega$ has a diffeomorphism--invariant cyclic vector $\Omega$ with $\omega(a)=\langle \Omega,\pi(a)\Omega\rangle$, which turns out to be an eigenstate of spatial geometry, which represents the completely degenerate geometry. The GNS--Hilbert space has a basis spanned by spin--network functions, which are eigenstates of quantum geometry and are mutually orthogonal.
  \item An important feature of the GNS--functional $\omega$ is that it is invariant under the pull--back action of spatial diffeomorphisms. This allows one to implement the pull--back action of spatial diffeomorphisms as unitary transformations, which is one of the great successes of kinematic Loop Quantization.
\end{enumerate}

A similar approach can be used to construct ``polymer quantization'' of geometrodynamics with a vacuum state that is invariant under the pull--back action of spatial diffeomorphisms and spatial conformal transformations. The operator algebra is constructed as follows:\\
For each bounded functional of the metric $F[g]$, we define an operator $\hat{F}[g]$ and for each real symmetric 2-tensor field $f_{ab}$ we define an operator $W(f)$. The product is defined through
\begin{equation}
 \widehat{F_1}[g]\widehat{F_2}[g]=\widehat{F_1f_2}[g],\,\,\,\,\,W(f_1)W(f_2)=W(f_1+f_2)
\end{equation}
and the canonical commutation relations
\begin{equation}
 W[-f]\hat{F}[g]W[f]=\hat{F}[g+f].
\end{equation}
We require that the unit functional $F_e:g\mapsto 1$ is quantized as the unit operator and that $W[f\equiv0]$ is quantized as the unit operator. The involution is defined through
\begin{equation}
 \hat F[g]^*=\hat{\overline{F}}[g],\,\,\,\,\,\,W[f]^*=W[-f].
\end{equation}
This defines an operator algebra $\mathfrak A_o$ for geometrodynamics, in which the $W[f]$ are interpreted as the Weyl operators $W[f]=\exp({i\int d^3x\,f_{ab}\hat\pi^{ab}})$, as the finite sums or ordered pairs
\begin{equation}
 a=\sum_{i=1}^n \hat F_i[g] W[f_i].
\end{equation}
On this algebra, we define the linear functional 
\begin{equation}
 \omega(a)=\sum_{i=1}^n \delta[f_i]F[g\equiv 0],
\end{equation}
where $\delta[f_i]=1$ when $f_i\equiv 0$ and vanishes otherwise. Direct calculation assuming w/o.l.o.g. that each $f_i$ occurs at most in one summand in $a$ shows
\begin{equation}
 \omega(a^*a)=\sum_i||F_i[f_i]||^2\ge 0,
\end{equation}
so $\omega$ can be used to perform the GNS--construction. The GNS--representation $\pi:\mathfrak A_o \to B(\mathcal H)$ on the GNS--Hilbert space $\mathcal H$ has a cyclic vector $\Omega$ which is an eigenstate of spatial geometry associated with the completely degenerate geometry and $\omega(a)=\langle \Omega,\pi(a)\Omega\rangle$. A basis of $\mathcal H$ is given by $\pi(W(f))\Omega$, which are eigenstates of spatial geometry with metric $f_{ab}$ and are mutually orthogonal. Moreover, the vacuum is invariant under spatial diffeomorphisms and spatial conformal transformations, which is the prerequisite to implement the pull-backs under diffeomorphisms and conformal transformations as unitary operators. 

This quantization program is not completed. But if it was successfully completed, one would have a Hilbert--space with unitary representation of the diffeomorphisms and conformal transformations and one can attack the problem of quantizing the dynamics. This problem may be aided by the observation that we made before: Given the conformal gauge--fixing condition described in section \ref{sec:usefulGauge}, we found an explicit expression for the Hamiltonian. The aiding idea is to first quantize this simple expression and to implement conformal invariance subsequently, e.g. through group averaging.

\subsection{Local effective Field Theories (Modified SD)}

We saw that all current approaches to quantize Shape Dynamics are incomplete and it may turn out that they will never be completed. However, even in absence of a fundamental quantum theory, one can consider effective modified gravity theories motivated by Shape Dynamics. A particular source for inspiration is the derivative expansion of Shape Dynamics, which is based on the observation that the solution $\Omega$ to the Lichnerowicz-York equation $\Omega^7(8\Delta-R)\Omega=T\Omega^{12}-M$ can be obtained analytically if the initial data satisfies
\begin{equation}\label{equ:homogeneityCondition}
 \frac{\sigma^a_b\sigma^b_a}{|g|}(x)=const.,\,\,\,\,\,\,R[g,x)=0.
\end{equation}
The second condition can actually be weakened to $R[g,x)=const.$, but I will abstain from this generality for now, because the scalar curvature contains spatial derivatives and I seek an expansion in the number of spatial derivatives. To do this, I write the Lichnerowicz-York equation as
\begin{equation}\label{equ:LYDerivative}
 \epsilon\, \Omega^7(8\Delta-R)\Omega=T\Omega^{12}-M
\end{equation}
where I introduced the counting parameter $\sqrt{\epsilon}$ which keeps track of the number spatial derivatives. Next, I expand $\Omega$ as a series in $\epsilon$ around the analytic solution $\Omega_o$:
\begin{equation}\label{equ:OmegaAnsatz}
 \Omega=\Omega_o +\sum_{n=1}^\infty \epsilon^n \Omega_n.
\end{equation}
This expansion obviously truncates after $\Omega_o$ if the initial data satisfies the homogeneity conditions in equation (\ref{equ:homogeneityCondition}). One can thus see the derivative expansion based on equation (\ref{equ:LYDerivative}) and the ansatz (\ref{equ:OmegaAnsatz}) as a perturbation around data that satisfies the homogeneity conditions of (\ref{equ:homogeneityCondition}).

Inserting the ansatz (\ref{equ:OmegaAnsatz}) into equation (\ref{equ:LYDerivative}) allows me to solve the order of $\epsilon^n$ for $\Omega_n$, because $\Omega_n$ appears only on the RHS and only in the term $n \Omega_o^{n-1} \Omega_n$. More precisely, by comparing the coefficients of $\epsilon^n$ on the left- and right hand side, I find
\begin{equation}
 \begin{array}{rcl}
   0&=& T\,\Omega_o^{12}-M\\
   \Omega_o^7(8\Delta-R)\Omega_o&=&12\,T\,\Omega_o^{11}\,\Omega_1\\
   LHS_n(\Omega_o,...,\Omega_{-1},8\Delta-R)&=&12\,T\,\Omega_o^{11}+RHS_n(\Omega_o,...,\Omega_{n-1},T),
 \end{array}
\end{equation}
where $LHS_n$ and $RHS_n$ are obtained by comparing the coefficients of $\epsilon^n$. The solution to this tower of recursive equations is
\begin{equation}
 \begin{array}{rcl}
   \Omega_o&=&\left(\frac{M}{T}\right)^\frac{1}{12}\\
   \Omega_1&=&(12T\Omega_o^4)^{-1}(8\Delta-R)\Omega_o\\
   \Omega_n&=&\frac{LHS_n-RHS_n}{12T\Omega_o^{11}},
 \end{array}
\end{equation}
so $\Omega_1=\frac{1}{12 T} \left(\frac{T}{M}\right)^{\frac{1}{3}}(8\Delta-R)\left(\frac{M}{T}\right)^\frac{1}{12}$.

This allows me to write the Hamiltonian $H=\int d^3x \sqrt{|h|}\Omega^6$ as a series in $\epsilon$; the first orders are
\begin{equation}
 \begin{array}{rcl}
  H&=&\int d^3x \sqrt{|g|}\left(\Omega_o^6+\epsilon \,6\Omega_o^5\Omega_1+\epsilon^2\left(15 \Omega_o^4\Omega^2_1+6\Omega_o^5\Omega_2\right)+\mathcal O(\epsilon^3)\right).
 \end{array}
\end{equation}
For the zeroth and first order of the Hamiltonian, I obtain the explicit expression as
\begin{equation}\label{equ:Hamiltonian1stOrder}
 H=\int d^3x \sqrt{|g|}\left(\sqrt{\frac{M}{T}}+\epsilon\,\frac{1}{2T}\,\left(\frac{M}{T}\right)^{\frac{1}{12}}(8\Delta-R)\left(\frac{M}{T}\right)^{\frac{1}{12}}\right)+\mathcal O(\epsilon^2)
\end{equation}
It is important to notice that, expect for the first order, no finite truncation of the derivative expansion is strictly conformally invariant. The truncations are rather conformally invariant up to spatial derivatives beyond the truncation. It is however not difficult to look at the truncation in equation (\ref{equ:Hamiltonian1stOrder}) and change the second order term in a way that makes it conformally invariant by replacing it with $f(T) CS(\Gamma[g])$, where $CS(\Gamma)$ denotes the Chern-Simons density functional of the Christoffel connection and $f$ an arbitrary function, this leads to a {\it local} modified Shape Dynamics Hamiltonian
\begin{equation}\label{equ:effHam}
 H_{mod}=\int d^3x\left(\sqrt{\frac{M}{T}}+f(T)CS(\Gamma[g])+...\right),
\end{equation}
where $...$ stands for further modifications. An interesting feature of $H_{mod}$ is that it coincides with the un--modified Shape Dynamics Hamiltonian when (\ref{equ:homogeneityCondition}) is satisfied and that the potential $CS(\Gamma[g])$ vanishes when $g_{ab}$ is conformally flat. These two features make the modifications hard to detect experimentally. 

\subsection{Applications of SD}

The conformal symmetry of Shape Dynamics presents a new perspective for some problems in gravity.

\subsubsection{Problem of Time in SD}

An important problem in canonical quantum gravity is the problem of time (see e.g. \cite{Anderson:2011jb} for an extensive review). This is due to the fact that time evolution and gauge--invariance are entangled in the ADM formulation of gravity. This entanglement is encoded in the ADM constraint algebra, which I repeat here
\begin{equation}
  \begin{array}{rcl}
    \{H(v_1),H(v_2)\}&=&H([v_1,v_2])\\
    \{H(v),S(N)\}&=&S(\mathcal L_v N)\\
    \{S(N_1),S(N_2)\}&=&H(N_1\nabla N_2-N_2\nabla N_1),
  \end{array}
\end{equation}
and the fact that the scalar constraints $S(N)$ play a dual role: serving on the one hand a generators of refoliations (gauge transformations) and on the other hand as generators of dynamics. This is complicated by the fact that one can not simply choose a single scalar constraint, say $S(N\equiv 1)$, and reinterpret it as a physical Hamiltonian rather than a constraint, because closure of the constraint algebra combined with linearity and $\{H(v),S(N)\}=S(\mathcal L_v N)$ implies that all scalar $S(N)$ are constraints. This is a serious problem in canonical quantum gravity, because Dirac quantization requires the wave function of the universe to be annihilated by all constraints, so no time--evolution can occur in the quantum theory. 

Shape Dynamics cleanly disentangles gauge--invariance from evolution: The constraints generate spatial diffeomorphisms and conformal transformations and the system is evolved with a physical Hamiltonian. Dirac quantization of the Shape Dynamics description of gravity thus requires that the wave function of the universe is annihilated by the generators of spatial diffeomorphisms and conformal transformations and the wave function evolves with a physical Hamiltonian. The problem of time is avoided; the price for this is the complicated form of the Shape Dynamics Hamiltonian.

\subsubsection{Shape Dynamics interpretation of holographic renormalization}

The bulk-bulk equivalence of a gravitational theory (General Relativity) and a conformal theory (Shape Dynamics) is similar to familiar the bulk-boundary duality of the very familiar AdS/CFT framework of \cite{Maldacena:1997re,Witten:1998qj}. Some aspects of the AdS/CFT framework can be understood as a boundary manifestation of the bulk-bulk equivalence between General Relativity and Shape Dynamics using the notion of the duality established in \cite{Freidel:2008sh}.

{\it Asymptotic Equivalence:} The usual semiclassical holographic renormalization is based on the near boundary expansion of the general solution to Einstein's equations when particular asymptotic boundary conditions are imposed. A large number of the asymptotic boundary conditions (see e.g. \cite{deHaro:2000xn,Papadimitriou:2004ap} can be translated into ADM evolution with asymptotically homogeneous lapse $N$ and with asymptotically vanishing shift $\xi$ and with asymptotic homogeneity of mean extrinsic curvature $\frac{\pi}{\sqrt{|g|}}$ and with asymtotically homogeneous spatial scalar curvature $R$. This means in particular that the ADM evolution is asymptotically in CMC--gauge, where the ADM evolution equations coincide with the Shape Dynamics evolution equations, because homogeneity of $\frac{\pi}{\sqrt{|g|}}$ implies that the ADM data is CMC and the remaining boundary conditions imply the CMC--lapse fixing equation
\begin{equation}
 \left(\Delta - R \frac 1 4 \langle\pi\rangle^2\right)N=\textrm{const.},
\end{equation}
which is solved by a homogeneous lapse $N$ \cite{Gomes:2013uk}. The boundary conditions thus imply that the ADM evolution coincides with the Shape Dynamics evolution. One can thus view the asymptotic conformal symmetry as being the remnant of the of the bulk conformal symmetry of Shape Dynamics. \\
The homogeneous lapse evolution is however only asymptotically a CMC evolution. The homogeneous lapse evolution ceases to be CMC as soon as any inhomogeneities (e.g. in $R$) enter in the lapse fixing equation. This is why the bulk conformal symmetry of Shape Dynamics is not found in the ADM evolution with homogeneous lapse.

{\it Manifest Equivalence:} Strong gravity (this is a regime where all spatial derivatives can be neglected) is a neat example where the homogeneous lapse propagates the CMC condition. One can see the equivalence explicitly by comparing the ADM homogeneous lapse Hamiltonian
\begin{equation}
 S(N\equiv 1)=\int d^3x\left(\frac{\pi^{ab}G_{abcd}\pi^{cd}}{\sqrt{|g|}}+2\Lambda\sqrt{|g|}\right)\approx 0
\end{equation}
on the ADM constraint surface, i.e. the surface $S(x)=\langle S\rangle\sqrt{|g|(x)}$ where the inhomogeneous ADM constraints are satisfied, with the volume constraint form of the Shape Dynamics Hamiltonian
\begin{equation}
 V^{-1}\left(\int d^3x\sqrt{\pi^{ab}(g_{ac}g_{bd}-\frac 1 3 g_{ab}g_{cd})\pi^{cd}}\right)^2+\left(2\Lambda-\frac 1 6\langle\pi\rangle^2\right)V\approx 0.
\end{equation}
A different approach to see the manifest equivalence between Shape Dynamics evolution and the homogeneous lapse evolution is to look at the first orders in a large volume expansion of the volume constraint form of the Shape Dynamics Hamiltonian as in \cite{Gomes:2011dc}. This can be used to investigate the large volume regime in a spherical slicing of an asymptotically deSitter universe. These first terms of the asymptotic Shape Dynamics Hamiltonian are in this case
\begin{equation}
 0 \approx \left( 2\Lambda - \frac 1 6 \langle\pi\rangle^2 \right) - \left(\frac{V_0}{V}\right)^{2/3} \langle\tilde R\rangle + \left(\frac{V_0}{V}\right)^2 \int_\Sigma d^3x\frac{\sigma^a_b\sigma^b_a}{\sqrt{\tilde g}} + O\left((\frac{V}{V_0})^{-8/3}\right),
\end{equation}
where the quantities with tilde are evaluated in Yamabe gauge, i.e. the conformal factor is gauged to satisfy $R[e^{4\hat\phi};x)=\langle R\rangle$. Direct inspection shows that the leading orders of the large volume expansion coincide with the large volume expansion of the ADM Hamiltonian with homogeneous lapse. However, the sub--leading orders $O\left((\frac{V}{V_0})^{-8/3}\right)$ exhibit explicit differences, which again shows that the homogeneous lapse evolution in general ceases to be CMC in the bulk and hence implies that the ADM homogeneous lapse evolution ceases to manifestly coincide with the Shape Dynamics evolution. 

{\it Semiclassical Limit of Renormalization Evolution:} There is an interesting connection with holographic renormalization. We saw in section \ref{sec:TheorySpace} that exact renormalization group equations can be derived using a scale--dependent mass term, which gives each mode of wavelength longer than the renormalization scale $k$ an effective mass term of order $m^2=k^2$ while effectively vanishing for shorter wavelength modes. This becomes an effective mass term in the UV--limit of the renormalization group equation. A semicalssical approximation to the UV--limit of the renormalization group equation is thus effectively the dynamics of a quadratic kinetic term, such as the kinetic term in the ADM evolution with homogeneous--lapse. This provides a heuristic explanation why gravitational dynamics should encode the UV--limit of renormalization group flow if it where not for the indefinite sign of the DeWitt supermetric $G_{abcd}$, which spoils the interpretation of the ADM kinetic term as a 
renormalization group kernel $R_k$. This is where Shape Dynamics provides an important insight, because the trace is pure gauge in Shape Dynamics. The kinetic term is positive semidefinite for physical (i.e. trace--free) degrees of freedom and can thus be interpreted as a renormalization group kernel for all physical degrees of freedom. 

\subsubsection{Shape Dynamics perspective on Conformal Mode}

We just saw that the interpretation of gravitational dynamics as renormalization group flow requires the identification of the trace degrees of freedom as pure gauge, as it is in Shape Dynamics. This is only a special instant of the general statement that there is no ``wrong--sign'' problem in gravity. The Shape Dynamics description reveals that the kinetic term for the physical degrees of freedom is positive semidefinite. The wrong sign appears only in the pure gauge sector.

\section{Summary}

The basic observation that underlies the construction of Shape Dynamics is very simple and generic: the same physical system can be accurately described using many mathematical models. In particular gauge symmetries are not properties of the physical system, but of the description in terms of local degrees of freedom. This fact allows one to trade one set of gauge symmetries for another and in particular to trade the spacetime refoliation symmetry of the ADM description of General Relativity for the spatial conformal symmetry of Shape Dynamics. Gravity can thus be viewed as a dynamical theory of spatial conformal geometry. This shift from spacetime geometry to spatial conformal geometry opens new possibilities for quantum gravity. In this contribution I discussed in particular:
\begin{enumerate}
 \item Some results in quantum gravity may be reinterpreted in light of Shape Dynamics. I showed this on the example of causal dynamical triangulations of the torus in 2+1 dimensions: The spacetime interpretation of Monte--Carlo simulations of causal dynamical triangulations of the 2+1 torus shows that the shape degrees of freedom evolve in according to Einstein's equations, but the spatial volume does not. Hence, the Shape Dynamics interpretation of the simulations is that they are in agreement with Einstein's equations, while the spacetime interpretation finds a mismatch with Einstein's equations.
 \item The Shape Dynamics description of gravity introduces a new theory space for quantum gravity based on action functionals for the conformal 3--geometry. This theory space is the same as the theory space for low--energy Hor\v{r}ava-Lifshitz theory. The resulting effective gravity theories are however quite different, because the Shape Dynamics ontology assumes conformal symmetry also in the quantum theory through {\it modified} Ward--identities. The natural effective Shape Dynamics Hamiltonian is thus of the form (\ref{equ:effHam}).
 \item One of the most promising approaches to quantum gravity is the Loop Quantum Gravity program. An open problem in Loop Quantum Gravity is the construction of a physically acceptable physical Hilbert space for pure quantum gravity. The difficulty is due to the complicated structure of the Hamilton constraints. It turns out that the conformal constraints of Shape Dynamics are equally complicated. However, using the geometric understanding of conformal transformations, one can find a heuristic quantization strategy for the conformal constraints and is thus in principle able to construct a physical Hilbert for the Loop Quantization of Shape Dynamics. 
 \item An alternative quantization procedure, besides Loop-- and Fock--quantization, is polymer quantization in metric variables. Polymer quantization allows the direct implementation of spatial diffeomorphisms and spatial conformal transformations and thus allows one to construct the physical Hilbert space for pure gravity. The complicated structure of the Shape Dynamics Hamiltonian however obstructs efforts to directly quantize it. Instead, systematic searches for Hamiltonians with physically acceptable low--energy limit will have to be undertaken.
 \item Classical Shape Dynamics disentangles the problem of gauge invariance (diffeomorphism and conformal symmetry) form dynamics (Shape Dynamics Hamiltonian). One can thus Dirac quantize Shape dynamics without having to worry about the problem of time.
 \item The bulk--bulk equivalence between a theory of spacetime geometry (the standard description of General Relativity) and a conformal theory of spatial geometry (the Shape Dynamics Description of General Relativity) is very reminiscent of the bulk--boundary duality in the AdS/CFT framework. It turns out in this context, that some aspects of semiclassical holographic renormalization correspondence can be viewed as the consequence of the bulk--bulk equivalence. This is particularly clear in the large volume limit of asymptotically deSitter solutions.
\end{enumerate}
The applications of the Shape Dynamics description of gravity go of course beyond the topics discussed in this article, where I restricted myself to those topics where at least partial results are obtained. There are many interesting questions beyond the topics covered in this contribution. E.g. current research is concerned with the investigation of the question whether there global Shape Dynamics solutions, which are of course locally solutions to Einstein's equations, but which are not globally equivalent to solutions of Einstein's equations. This seems to be the case for the Einstein--Rosen bridge, which is a vacuum solution for Shape Dynamics, but not the spacetime description of General Relativity. 

\subsection*{Acknowledgements}

This work was supported by the Natural Sciences and Engineering Research Council of Canada. I am very grateful for useful discussions with my closest collaborator, Henrique Gomes, and many stimulating discussions with Viqar Husain. Furthermore, I am very grateful to Julian Barbour, Sean Gryb, Flavio Mercati, Timothy Budd and Lee Smolin for various collaborations on the Shape Dynamics formulation of General Relativity.

\bibliographystyle{plain}
\bibliography{tim}

\end{document}